\documentclass[%
 aip,
 amsmath,amssymb,
reprint,%
]{revtex4-1}

\usepackage{graphicx}% Include figure files
\usepackage{dcolumn}% Align table columns on decimal point
\usepackage{bm}% bold math
\usepackage{caption}
\usepackage{subcaption}
\usepackage{xcolor}
\usepackage[utf8]{inputenc}
\usepackage[T1]{fontenc}
\usepackage{mathptmx}
\begin{document}

\preprint{AIP/APL}

\title{Spectral control of high order harmonics through non-linear propagation effects}% \\textcolor{red}{Spectral control of high order harmonics through non-linear propagation effects in Si and ZnO}}

% Force line breaks with \\
\author{M. Hussain}
\email{mukhtar.hussain@tecnico.ulisboa.pt}
\affiliation{LIDYL, CEA, CNRS, Université Paris-Saclay, CEA Saclay 91191, Gif-sur-Yvette, France.%\\This line break forced with \textbackslash\textbackslash
}%
 \affiliation{GoLP/Instituto de Plasmas e Fusão Nuclear-Laboratório Associado, Instituto Superior Técnico, Universidade de Lisboa, 1049-001 Lisboa, Portugal.}
 
\author{S. Kaassamani}%
 %\email{Second.Author@institution.edu.}
\affiliation{LIDYL, CEA, CNRS, Université Paris-Saclay, CEA Saclay 91191, Gif-sur-Yvette, France.%\\This line break forced with \textbackslash\textbackslash
}%
\author{T. Auguste}
\affiliation{LIDYL, CEA, CNRS, Université Paris-Saclay, CEA Saclay 91191, Gif-sur-Yvette, France.%\\This line break forced with \textbackslash\textbackslash
}%
\author{W. Boutu}%
\affiliation{LIDYL, CEA, CNRS, Université Paris-Saclay, CEA Saclay 91191, Gif-sur-Yvette, France.%\\This line break forced with \textbackslash\textbackslash
}%
\author{D. Gauthier}%
 %\email{Second.Author@institution.edu.}
\affiliation{LIDYL, CEA, CNRS, Université Paris-Saclay, CEA Saclay 91191, Gif-sur-Yvette, France.%\\This line break forced with \textbackslash\textbackslash
}%
\author{M. Kholodtsova}%
 %\email{Second.Author@institution.edu.}
\affiliation{LIDYL, CEA, CNRS, Université Paris-Saclay, CEA Saclay 91191, Gif-sur-Yvette, France.%\\This line break forced with \textbackslash\textbackslash
}%
\author{J-T. Gomes}%
 %\email{Second.Author@institution.edu.}
\affiliation{Novae, 15 Rue Sismondi, 87000 Limoges, France.}%
\author{L. Lavoute}%
 %\email{Second.Author@institution.edu.}
\affiliation{Novae, 15 Rue Sismondi, 87000 Limoges, France.}%
\author{D. Gaponov}%
 %\email{Second.Author@institution.edu.}
\affiliation{Novae, 15 Rue Sismondi, 87000 Limoges, France.}%
\author{N. Ducros}%
 %\email{Second.Author@institution.edu.}
\affiliation{Novae, 15 Rue Sismondi, 87000 Limoges, France.}%
\author{S. Fevrier}%
 %\email{Second.Author@institution.edu.}
\affiliation{Novae, 15 Rue Sismondi, 87000 Limoges, France.}%
\affiliation{Univ. Limoges, CNRS, XLIM, UMR 7252, 87000 Limoges, France.}%

\author{R. Nicolas}%
 %\email{Second.Author@institution.edu.}
\affiliation{LIDYL, CEA, CNRS, Université Paris-Saclay, CEA Saclay 91191, Gif-sur-Yvette, France.}%
\affiliation{Natural Sciences Department, Lebanese American University, Beirut, Lebanon.}%

\author{T. Imran}
\affiliation{%
Research Laboratory of Lasers (RLL)-Group of Laser Development (GoLD), Department of Physics, COMSATS University Islamabad, Park Road 45550, Islamabad, Pakistan.%\\This line break forced% with \\
}%

\author{G. O. Williams}
\affiliation{%
GoLP/Instituto de Plasmas e Fusão Nuclear-Laboratório Associado, Instituto Superior Técnico, Universidade de Lisboa, 1049-001 Lisboa, Portugal.%\\This line break forced% with \\
}
\author{M. Fajardo}
\affiliation{%
GoLP/Instituto de Plasmas e Fusão Nuclear-Laboratório Associado, Instituto Superior Técnico, Universidade de Lisboa, 1049-001 Lisboa, Portugal.%\\This line break forced% with \\
}%
\author{H. Merdji}%
 \email{hamed.merdji@cea.fr}
\affiliation{LIDYL, CEA, CNRS, Université Paris-Saclay, CEA Saclay 91191, Gif-sur-Yvette, France.%\\This line break forced with \textbackslash\textbackslash
}%

\date{\today}% It is always \today, today,

\begin{abstract}

High harmonic generation (HHG) in crystals has revealed a wealth of perspectives such as all-optical mapping of the electronic band structure, ultrafast quantum information and the creation of  novel all-solid-state attosecond sources. Significant efforts have been made to understand the microscopic aspects of HHG in crystals, whereas the macroscopic effects, such as non-linear propagation effects of the driving pulse inside the dense solid media and its impact on the HHG process is often overlooked. In this work, we study macroscopic effects by comparing two materials with distinct optical properties, silicon (Si) and zinc oxide (ZnO). By scanning the focal position of 85 fs, 2.123 $\mu$m wavelength pulses inside the crystals (Z-scan) we reveal spectral shifts in the generated harmonics. We interpret the overall blueshift of the emitted harmonic spectrum as an imprint of the driving field spectral modulation occurring during the propagation inside the crystal. This is supported with numerical simulations. This study demonstrates that through manipulation of the fundamental driving field through non-linear propagation effects, precise control of the emitted HHG spectrum in solids can be realised. This method could offer a robust way to tailor HHG spectra for a range of spectroscopic applications.

\end{abstract}

\maketitle

\section{Introduction}

In the last decade, solid-state HHG has been studied in a wide variety of crystals and amorphous solids \cite{ghimire2011observation,you2017anisotropic,you2017high}. The physical picture of solid-state HHG is in its infancy and is the subject of ongoing discussion  \cite{golde2008high,higuchi2014strong,hawkins2015effect, wu2016multilevel, tancogne2017impact, floss2018ab}. A better understanding of the physical process of solid-state HHG will not only lead to improved secondary sources, but to an ability to map the nonlinear properties and possibly allow all-optical mapping of the electronic structure of solids \cite{marangos2011high,schubert2014sub,hohenleutner2015real,luu2015extreme,vampa2015all,ghimire2018high}. HHG in solids must be realised in materials transparent to the driving laser pulse wavelength, which is generally in the near  to mid infra-red region. Therefore, insulating and semi-conducting solids are the materials of choice. The HHG process in solids shares some similarities with HHG in gases, at least conceptually \cite{vampa2015all,vampa2017merge}. The first step is an inter-band transition in the material that promotes a charge carrier to a conduction band. Once there, this electron is driven by the fundamental field, and can emit harmonics due to oscillations within a band (intra-band), or excitation and relaxation between bands (inter-band) \cite{vampa2015semiclassical,ghimire2018high}.

The initial inter-band excitation requires tunnelling or a multi-photon absorption process, and the rate of free carrier generation is therefore inversely proportional to the band gap. Apart from the microscopic physics of carrier generation and motion, the properties of the HHG emission can be influence by the propagation of the electric fields inside the crystal. This macroscopic propagation is nonlinear for the driving field and linear for the harmonics that have a much lower intensity. The emitted HHG spectrum can reflect the available electronic states of the material. This makes the HHG spectrum an intricate fingerprint of the complex interplay of microscopic and macroscopic processes.
For example, HHG in ZnO \cite{ghimire2011observation,gholam2017high, gholam2018high} has been shown to reveal both the inter and intra-band mechanisms coupled with the non-linear response of the crystal. Similarly, harmonics have been generated in the most widely used material in electronics, silicon (Si), at 2.1 $\mathrm{\mu m}$ driving wavelength \cite{vampa2016plasmonic,vampa2017plasmon} and used to disentangle the surface and bulk contributions \cite{vampa2019disentangling}. HHG in Si and ZnO can be controlled and monitored by manipulating the crystal  surfaces \cite{sivis2017tailored,franz2019all}. For example, spatial shaping of the crystal surface has been used to create HHG beams that carry an orbital angular momentum (OAM) \cite{gauthier2019orbital} without the need of expensive device for spatial phase control. These generated OAM beams have many applications varying from the manipulation of nanoparticles to quantum cryptography \cite{torres2011twisted}.

The propagation of the intense driving field has been previously linked with redshifts in the HHG spectrum in plasmas and gases \cite{brandi2006spectral,bian2013spectral,du2015nonadiabatic}. This effect is attributed to the delayed emission of HHG from excited states or resonant states \cite{bian2011nonadiabatic} or dislocation of molecules \cite{bian2014probing}. In solids, numerically computed HHG spectra have shown redshifts in the "higher plateaus" due to higher band transitions \cite{jia2017nonadiabatic}. In sapphire, intensity-dependent redshifts and blueshifts were observed in the spectrum of harmonics and attributed to the long and short electron/hole trajectories, respectively \cite{kim2019spectral}. These studies showed the electronic effects on the HHG process, while the macroscopic or propagation effects of the driving field on the spectral profile of harmonics hasn't been fully explored.

In the present study, we investigate the impact of the non-linear macroscopic beam propagation in the HHG process in Si and ZnO. We have generated 3rd (H3), 5th (H5) and 7th (H7) order-harmonics in Si and ZnO crystals at 2.123 $\mathrm{\mu m}$ driving central wavelength in the sub-TW/cm\textsuperscript{2} intensity regime. The spectral and spatial profiles of these high order harmonics are measured. This letter is structured as follows; the experimental setup to generate HHG and to observe the propagation effects of the driving field is described in section \ref{sec:exp}. HHG in Si and ZnO, and their spectral shift are reported in sections \ref{subsection:Spectral and spatial} and \ref{ZnO:spectral_spatial_Z}, respectively. Section \ref{subsection:modeling} presents a discussion of the experimental results linked with numerical propagation of a focused driving laser field in Si. Finally, the conclusion is reported in section \ref{sec:conclusion}.

\section{Experimental setup}
\label{sec:exp}

\begin{figure}
\centering
\includegraphics[width=7.75
cm, height=6 cm]{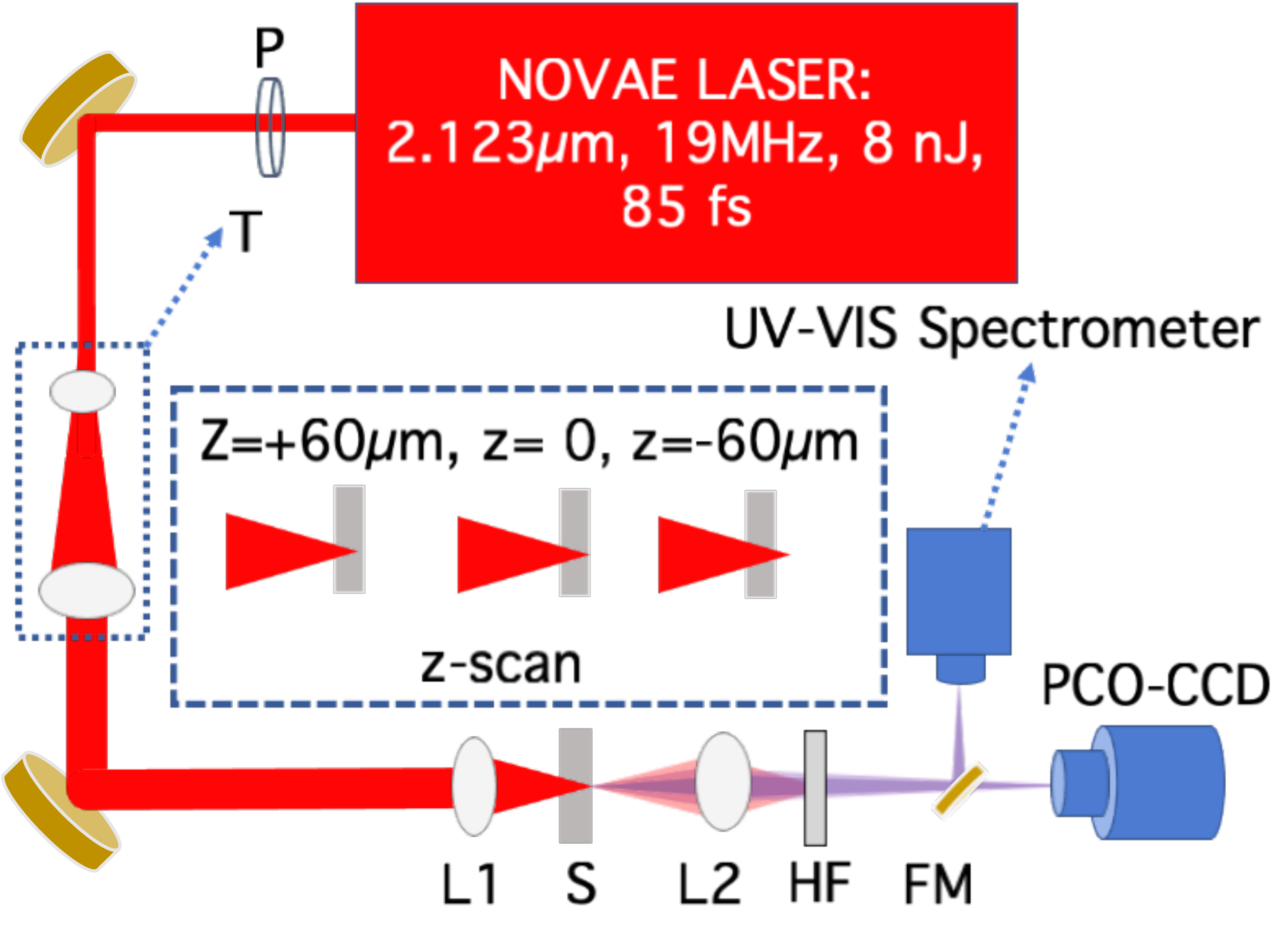}
	\caption{Schematic representation of the experimental setup to generate high order harmonics (H3, H5 and H7) in the 300 $\mathrm{\mu m}$ thick silicon (Si) and 200 $\mathrm{\mu m}$ thick-(100) oriented ZnO crystal. The convention for the Z-scan notation is explained in the inset.  Note that, P: polarizer, L: lens, T: telescope, S: sample, HF: harmonic filters, FM: flip mirror, CCD: charged coupled device.}
	\label{fig:Schematic}
\end{figure}

A schematic representation of the crystal HHG experimental setup is shown in Fig. \ref{fig:Schematic}. We have employed a fibre laser (NOVAE Company)  operating at 2.123 $\mathrm{\mu m}$ with a pulse energy of  $\sim$ 8 nJ and a repetition rate of 19 MHz. To enable a tighter focus and thus achieve higher intensities, the size of the laser beam is magnified by a factor 3.75 times using a telescope (T). The 85 fs duration pulses are focused in the crystals by a convex lens of 3 cm focal length up to a maximum peak intensity of  $\sim$ 0.38 TW/cm\textsuperscript{2} well below the damage threshold of crystals. The generated diverging harmonics are further focused by a convex lens of 15 cm focal length either on a UV-enhanced CCD camera or a UV-VIS spectrometer, for spatial and spectral beam characterization, respectively. Band-pass filters centred at each harmonic wavelength are used to separate and characterise the HHG properties. Using the optimized coupled spectrometer, Z-scans are performed to localize the longitudinal medium response as illustrated in the inset of figure \ref{fig:Schematic}.

\section{Results and discussions}
\subsection{HHG in Si and spectral shifting of harmonics}
\label{subsection:Spectral and spatial}
\begin{figure}
            \centering
             \begin{subfigure}{0.23\textwidth}
	            \includegraphics[width=4.2cm, height=4cm]{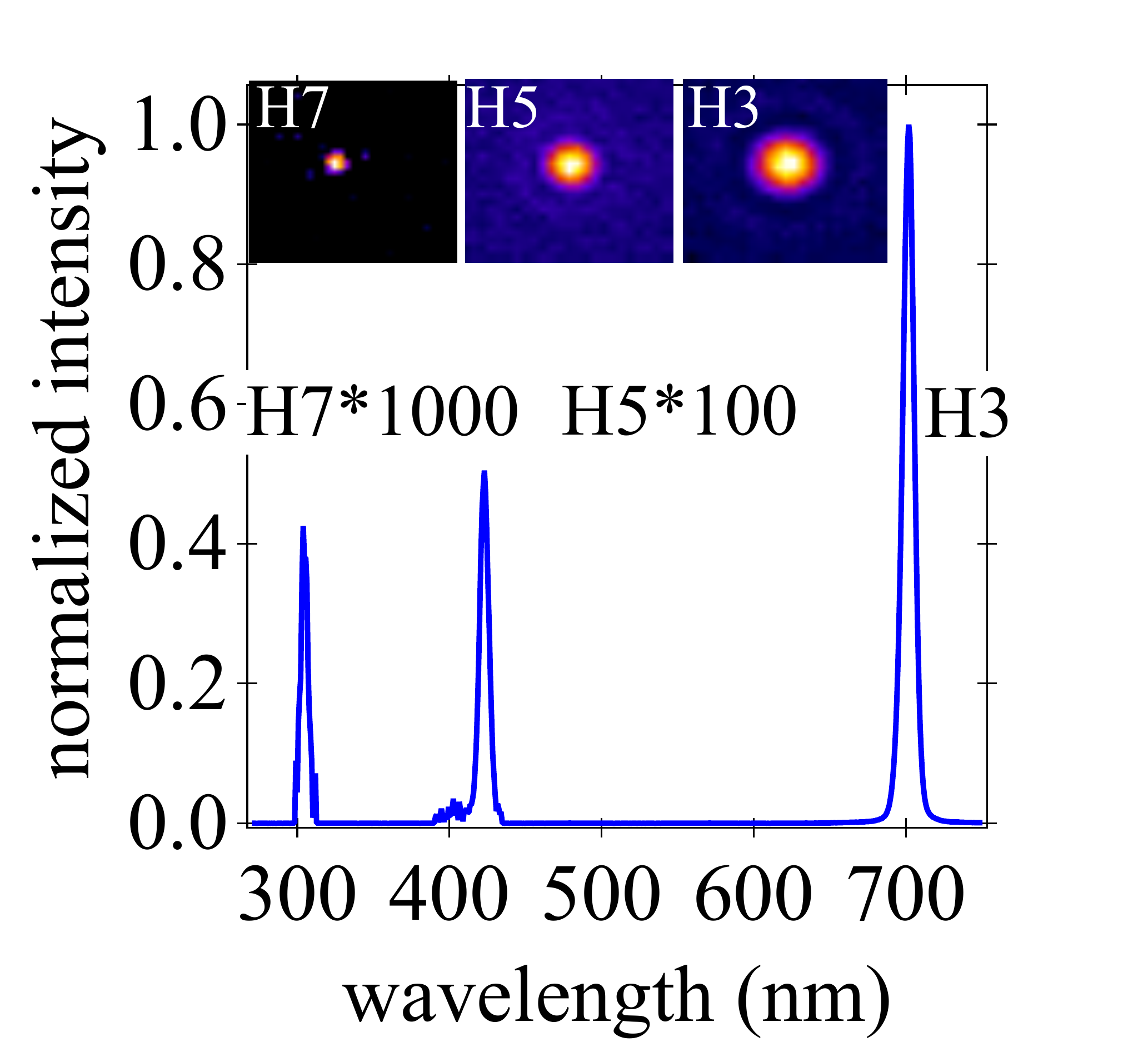}
		    	\caption{}
		    	\label{fig:Si_spectrum}
		     \end{subfigure}
             \centering
            \begin{subfigure}{0.23\textwidth}
	         \includegraphics[width=4.3cm, height=4.2cm]{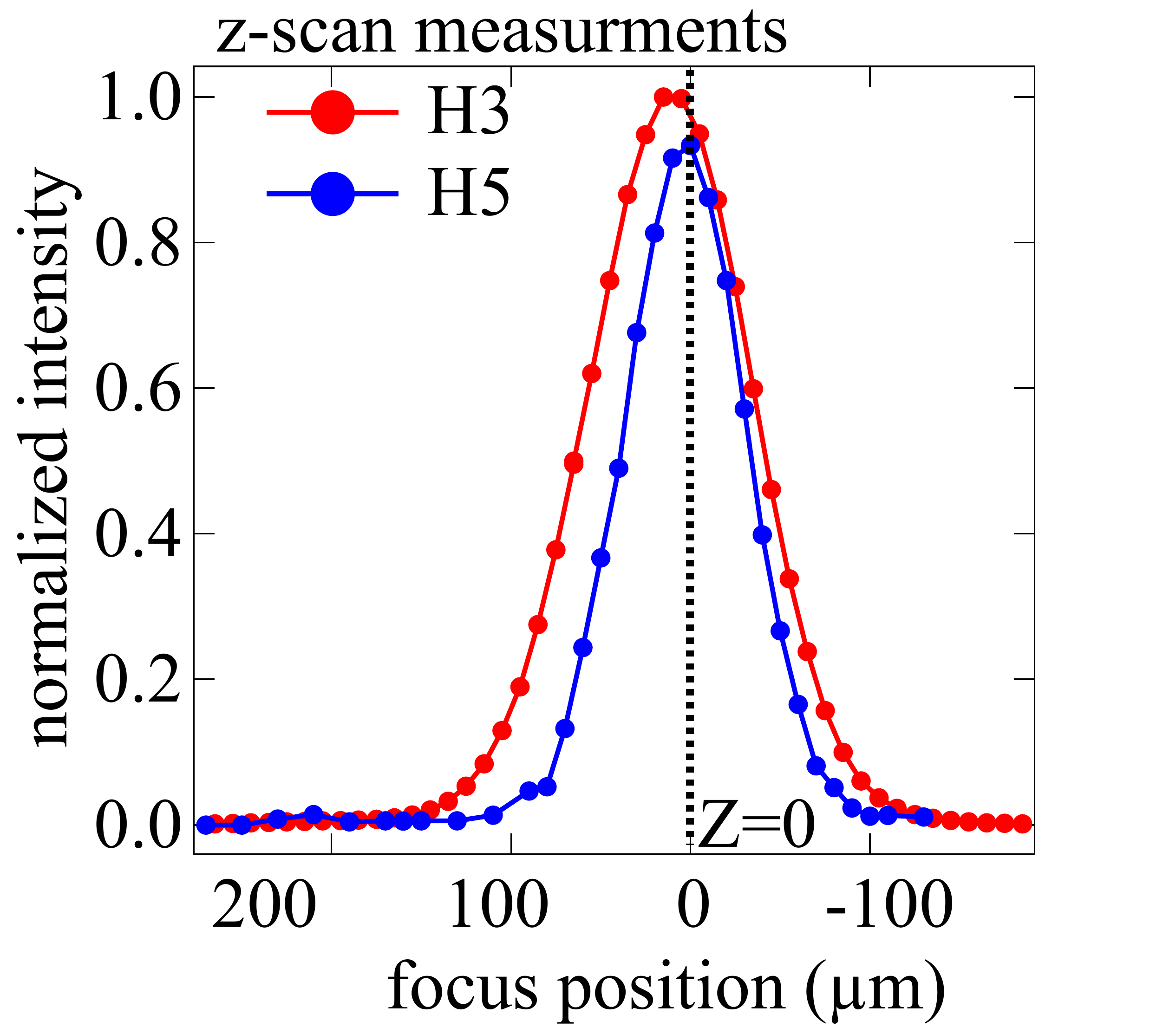}
	         \caption{}
	         \label{fig: Z scan}
	        \end{subfigure}
		    	\caption{ Spectral and spatial measurements of high order harmonics in a 300 $\mathrm{\mu m}$ thick Si crystal. The spectral and spatial profile of each harmonic measured separately with the corresponding harmonic filters for different acquisition time. (a) H5 multiplied by 100 and H7 by 1000 to make them visible in the combined graph. (b) Z-scan measurements of H3 (red dots) and H5 (blue dots) in the Si crystal.}
\end{figure}

 We have generated harmonics in a 300 $\mathrm{\mu m}$ thick, (100) oriented, Si crystal. The spectral and spatial profiles of H3, H5 and H7 harmonics are shown in Fig. \ref{fig:Si_spectrum}. The z-scans for H3 and H5 are recorded with steps of 10 $\mathrm{\mu m}$ as shown in Fig. \ref{fig: Z scan}. The signal of H7 is very weak and is therefore excluded from the Z-scan study. The yield of H3 and H5 reaches their maximum when the focus position of the driving field on the back surface of the Si crystal. This is attributed to the absorption of the harmonics generated at the front surface and inside the Si crystal.Compared to H3, the generation region of H5 is translated $\sim$ 5 $\mathrm{\mu m}$ towards the back surface due to its stronger absorption in the crystal (see Fig. \ref{fig: Z scan}). In parallel, we have measured the intensity-dependent spectral shifts of the harmonics generated in Si. 
 
\begin{figure}
    \centering
    \includegraphics[scale=0.3]{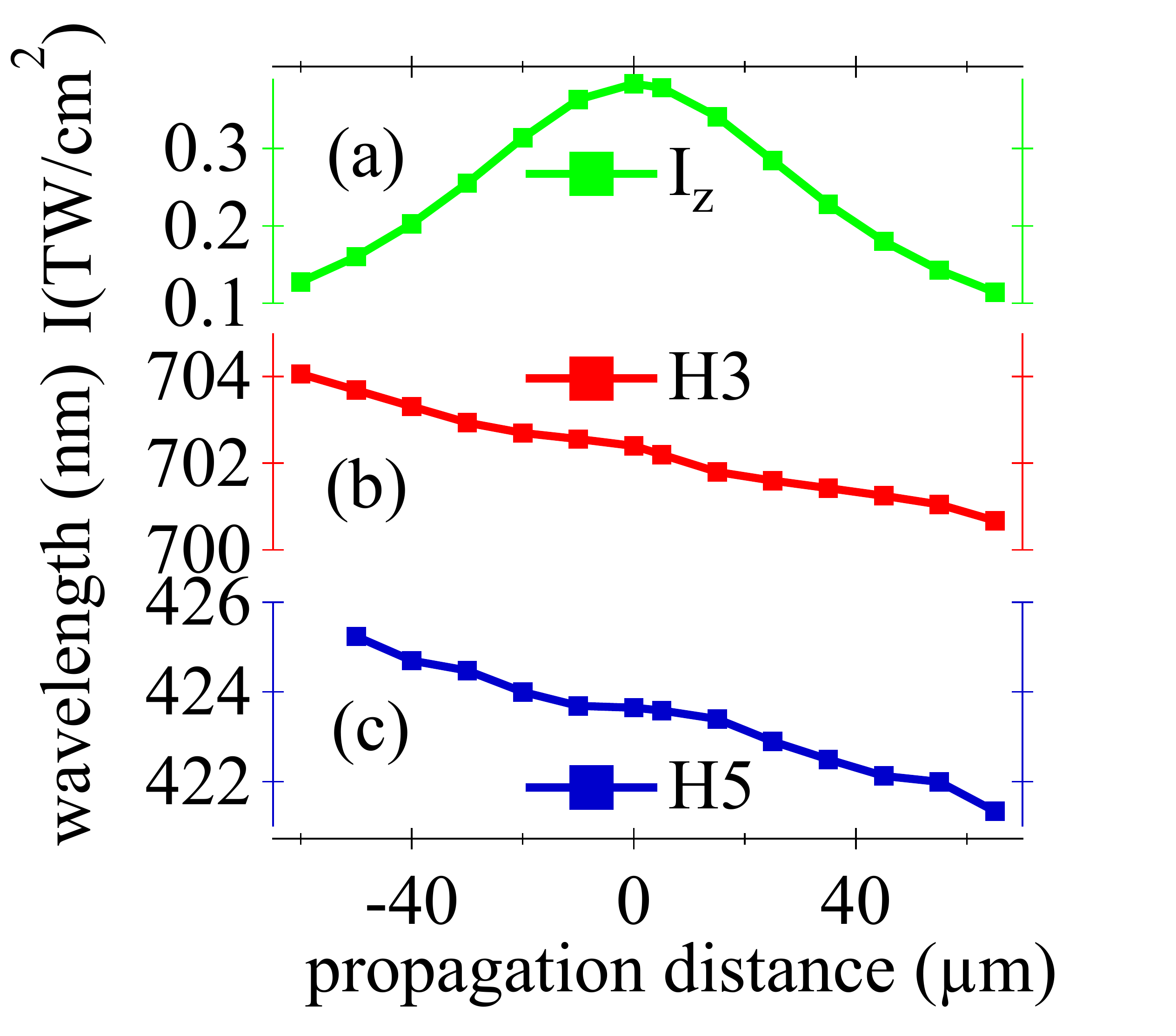}
    \caption{The intensity-dependent spectral blueshifts of the harmonic signal generated in the 300 $\mathrm{\mu m}$ thick Si crystal. (a) The driving intensity evolution ($I_z$) at the rear face of the crystal with the propagation distance "z". (b) spectral shift in H3 (c) spectral shift in H5.}
    \label{fig:Hshift_Si}
\end{figure}

 The driving intensity ($I_z$) at the surface of the crystal is estimated as $I_z$=$2P/\pi w_z^{2}$, where $w_z$ is the beam waist at z calculated using hyperbolic function \cite{bandres2004ince} by considering the non-linear refractive indices of crystals with a confocal parameter b=46 $\mathrm{\mu m}$ and $P$ is the peak power which is 0.05 MW in our case shown in Fig. \ref{fig:Hshift_Si}a. An intensity-dependent spectral shift of about 3.38 nm for H3 and 3.9 nm for H5 is observed in Si (see. Fig. \ref{fig:Hshift_Si}b and \ref{fig:Hshift_Si}c) show the intensity-dependent spectral shift of H3 and H5 generated from Si. As the laser focus is scanned from outside the crystal (at the position -60 $\mathrm{\mu m}$) to inside the bulk (at the position of +60 $\mathrm{\mu m}$) a spectral blue shift in H3 and H5 is observed. Indeed, when the laser focus is outside the crystal, the spectrum of H3 is centered at 704 nm, while that of H5 is centered at 425 nm. In this configuration, we expect that nonlinear propagation effects are minimal. However when the laser focus passes inside the crystal, the spectrum of H3 shifts to 701 nm, while that of H5 is shifts to 421 nm. In this case, the local field intensity is very high and triggers nonlinear effects.

\subsection{HHG in ZnO and spectral shifting of harmonics}
\label{ZnO:spectral_spatial_Z}

\begin{figure}
	        \centering
		    \begin{subfigure}{0.23\textwidth}
			    \includegraphics[width=4.3cm, height=4.2cm]{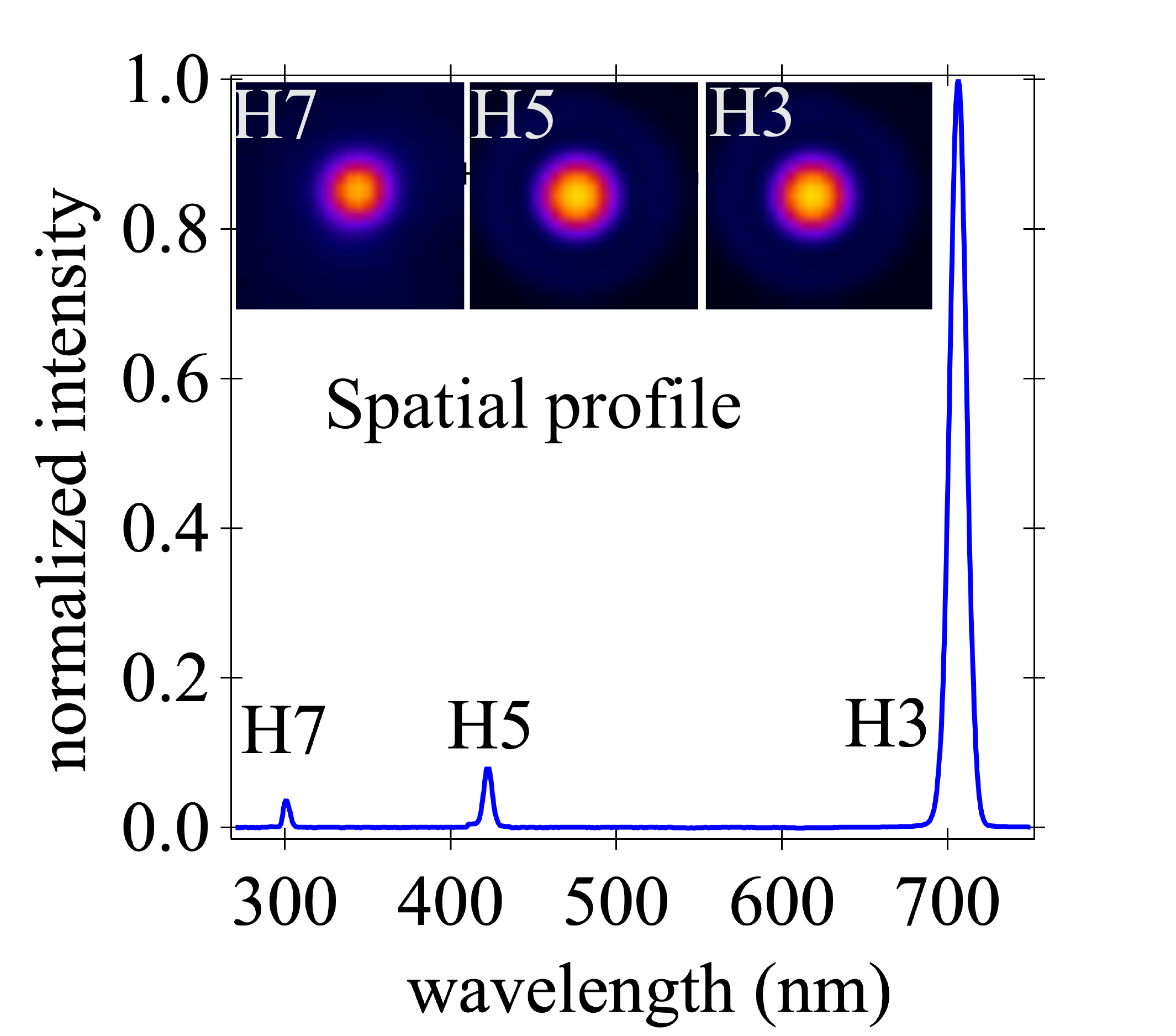}
			    \caption{}
			    \label{fig:ZnO_Spectrum}
		    \end{subfigure}
	       \centering
		    \begin{subfigure}{0.23\textwidth}
			    \includegraphics[width=4.5cm, height=4.5 cm]{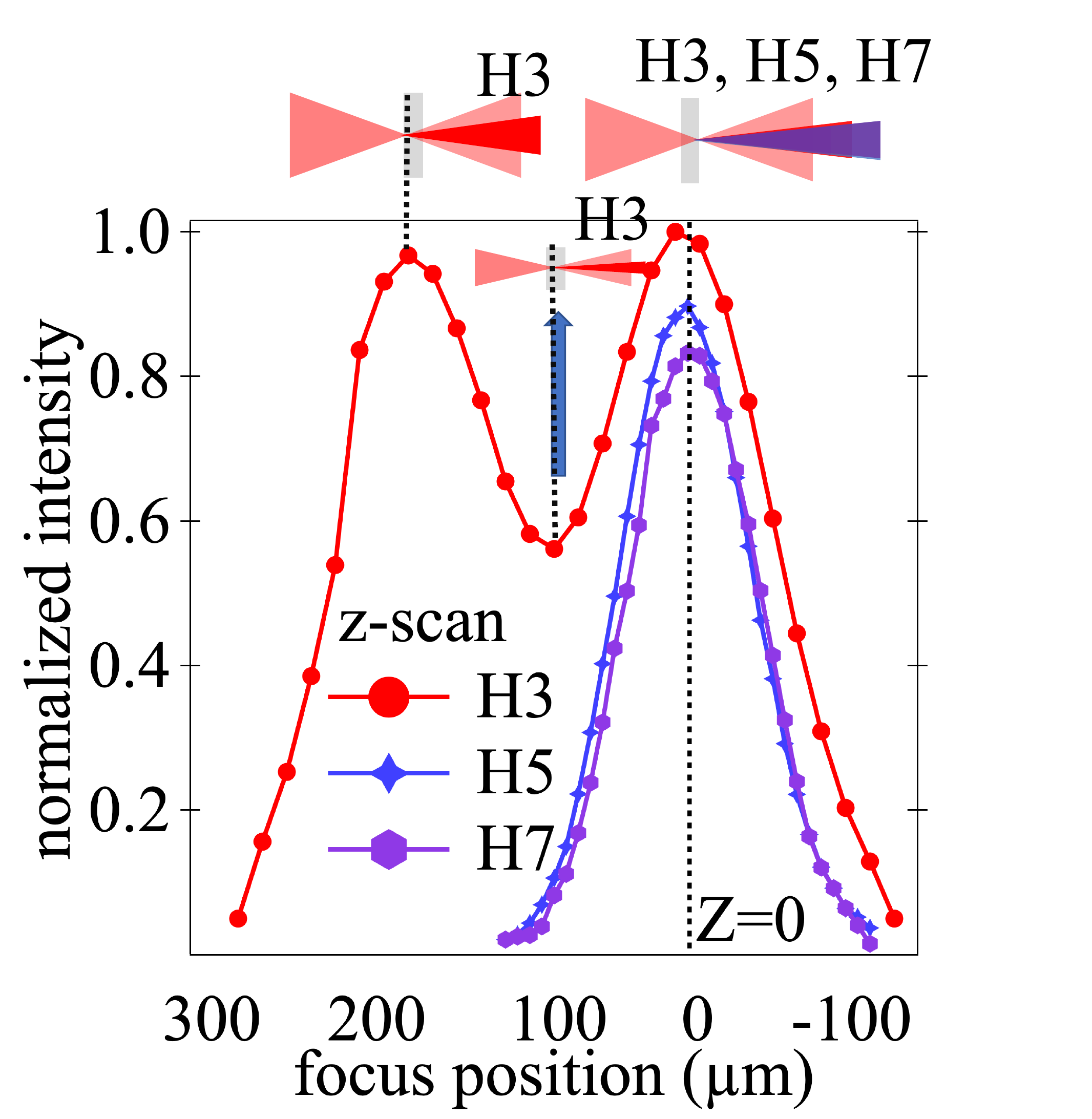}
			    \caption{}
			    \label{fig:ZnO_Z_scan}
		    \end{subfigure}
			\caption{(a) Spectrum of the harmonic emission in ZnO at z=0 for different acquisition time, with the harmonic beam profiles in insets, (b) Z-scan measurements of H3 (red dots), H5 (blue diamonds) and H7 (violet hexagons) order harmonics.} 
\end{figure}

The combined harmonic spectrum (H3, H5 and H7) generated in ZnO and the corresponding spatial profiles are shown in Fig.  \ref{fig:ZnO_Spectrum}. The measured harmonic signal (H3, H5, and H7) as a function of the focus position of the driving beam in the ZnO crystal is shown in Fig. \ref{fig:ZnO_Z_scan}. The Z-scan of H3 exhibits a double peak structure. The first peak corresponds to the generation from the front side of the crystal, while the second peak at z=0 corresponds to generation from the rear side of the crystal. The signal of H3 is reduced when the laser peak intensity is in the centre of the crystal. H3 generated before and after the focus point in the bulk contribute destructively when exiting the surface. As a result, the signal of H3 drops when the laser peak intensity is at the centre of the crystal while the signal of H3 is maximum when the laser is focused near the surfaces as shown in Fig. \ref{fig:ZnO_Z_scan}. We observe a single peak for H5 and H7 which shows the detected harmonics originate from the rear surface of the crystal. Indeed, the harmonics H5 and H7 generated from the front surface and bulk of the crystal are absorbed. As the harmonic order increases, the region of efficient harmonic generation moves closer to the back surface of the crystal.

\begin{figure}
		
		 \begin{subfigure}{0.23\textwidth}
			    \includegraphics[width=4.5cm, height=4.5cm]{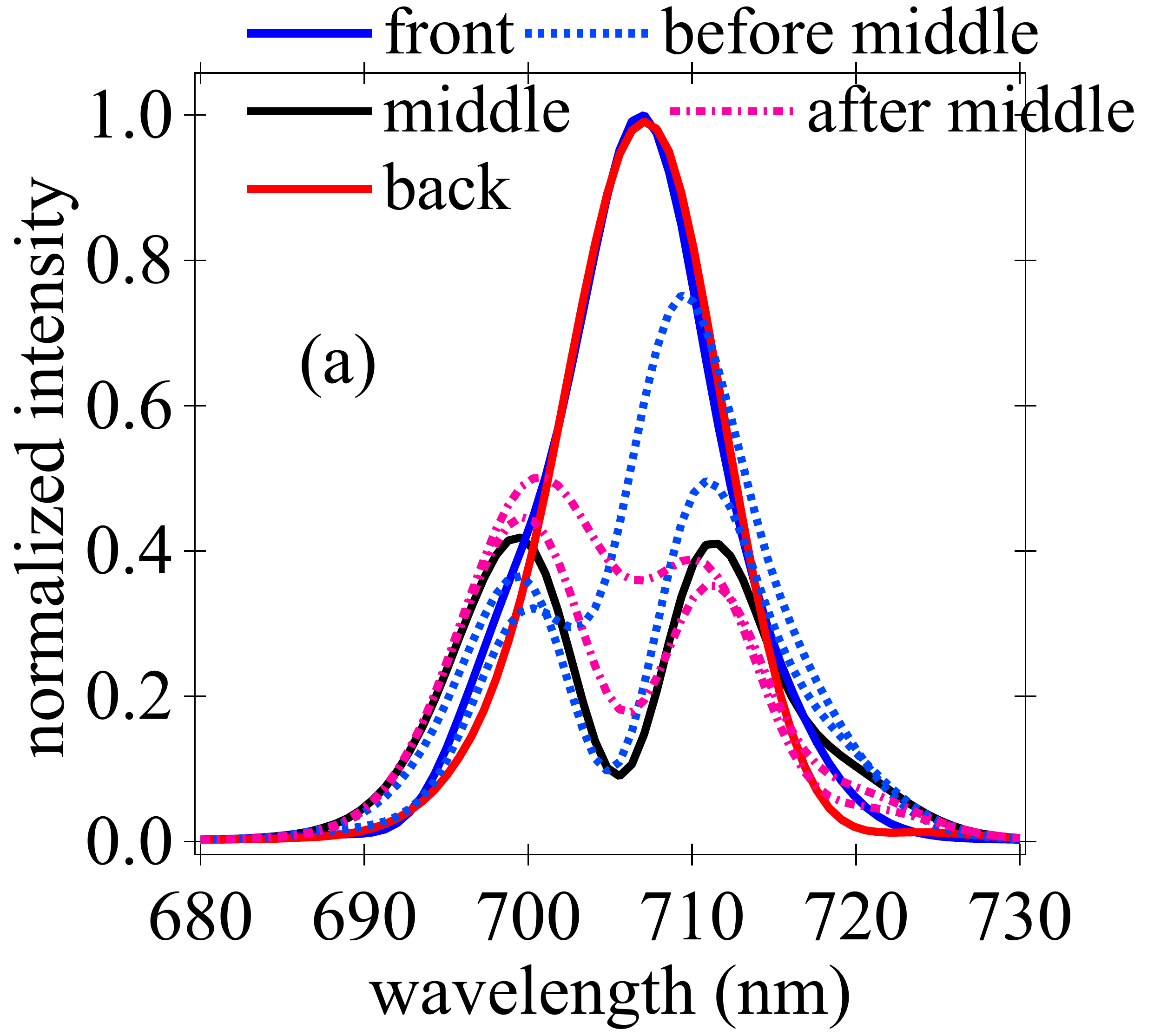}
			    %\caption{}
			    \label{fig:ZnO_H3_spectral}
		 \end{subfigure}
		   ~
	     \centering
		 \begin{subfigure}{0.23\textwidth}
			    \includegraphics[width=4.5cm, height=4.5cm]{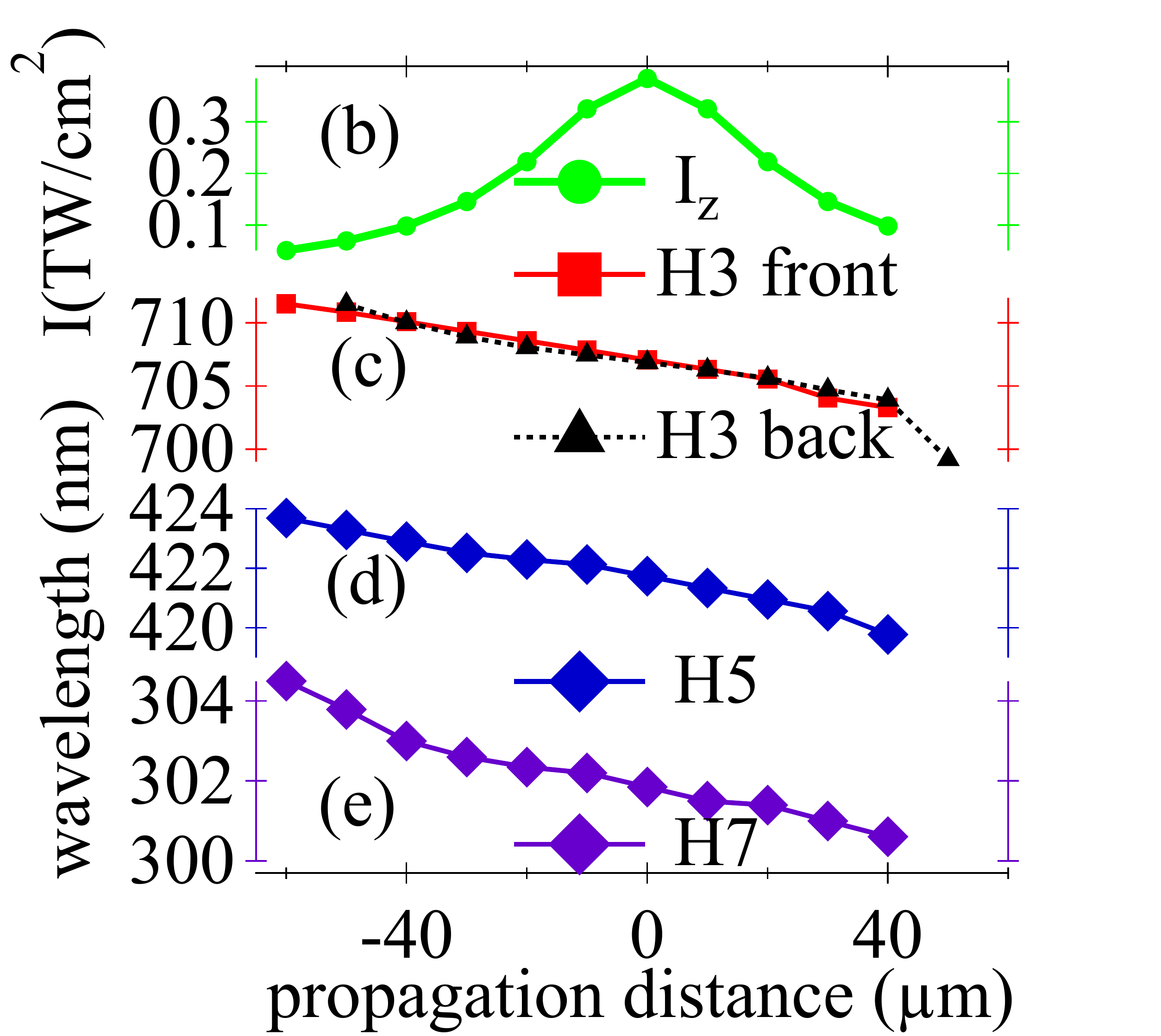}
			    %\caption{}
			    \label{fig:ZnO_H_shift}
		 \end{subfigure}
			\caption{(a) The intensity-dependent spectral shifting and splitting of H3 in ZnO (two curve statistics before and after the middle). (b) The evolution of the driving intensity ($I_z$) at the rear surface of crystal for different propagation distance (z) (green line). (c) H3 generated from the front surface (red data points) and at a back surface (triangle data points) of ZnO crystal. H3 exhibits a blueshift as the focus of the driving field translates towards the back and front surfaces. The z = 0 corresponds to the front surface for the red data points and the back surface of crystal for the triangle data points.  The Z-scan depth for HHG in terms of confocal parameters "b" is in the range of -1.3b to +0.9b, where b=46 $\mathrm{\mu m}$.  (d) spectral shift in H5, (e) spectral shift in H7.}
			\label{fig:ZnO_shift}
\end{figure}

We now turn to the corresponding spectral response of the harmonics. Blueshifts in the spectral profile of the harmonics are observed with the translation of the laser focus. Fig. \ref{fig:ZnO_shift}a shows the spectral response of H3 for different relative focus positions. A spectral modulation (shifting) and splitting of H3 peaks (double peak spectrum is observed) which are separated up to 12 nm when the focus of the driving field is in the middle of the ZnO crystal. The first peak of H3 begins to decrease while the second peak of H3 builds up with the translation of the driving focus towards the front surface. The intensity of H3 increases with the spectral blueshift while we move from outside the crystal towards the back surface of the crystal (from z = - 60 $\mathrm{\mu m}$ = - 1.30b to z = 0, triangle data points) and reached a maximum at z = 0 which corresponds to the rear surface of the ZnO crystal. 

When further translating the laser focus inside the crystal (z = + 40 $\mathrm{\mu m}$ = +0.9b), the intensity of H3 decreases. However, the width of the spectral shift increases even more. H3 generated from the front surface (red rectangle) and back surface (black triangle) of the crystal undergoes significant blueshifts (8.2 nm and 12.2 nm respectively) as shown in Fig. \ref{fig:ZnO_shift}c. This different spectral shift of H3 attributed to the different propagation distance of the intense driving beam at the front and back surface of crystal. We have observed spectral blueshifts in the spectrum of H5 and H7 as shown in Figs. \ref{fig:ZnO_shift}d and \ref{fig:ZnO_shift}e, respectively. There were no spectral modulations and splitting as observed for H3. The signals of H5 and H7 (Fig. \ref{fig:ZnO_Z_scan}) has increased when translating the laser peak intensity towards the rear surface. By further translating the focus of the driving field inside the crystal, the signal of H5 and H7 decreases. The total spectral shift across our measurement range is of about 3.91 nm for H5 and 3.89 nm for H7 and exhibits spectral blueshifts as shown in Figs. \ref{fig:ZnO_shift}d and \ref{fig:ZnO_shift}e, respectively.

\section{Modelling and Discussions}
\label{subsection:modeling}

\begin{figure}
	        \centering
	        \begin{subfigure}{0.22\textwidth}
	        \includegraphics[width=4.3 cm, height=4.2cm]{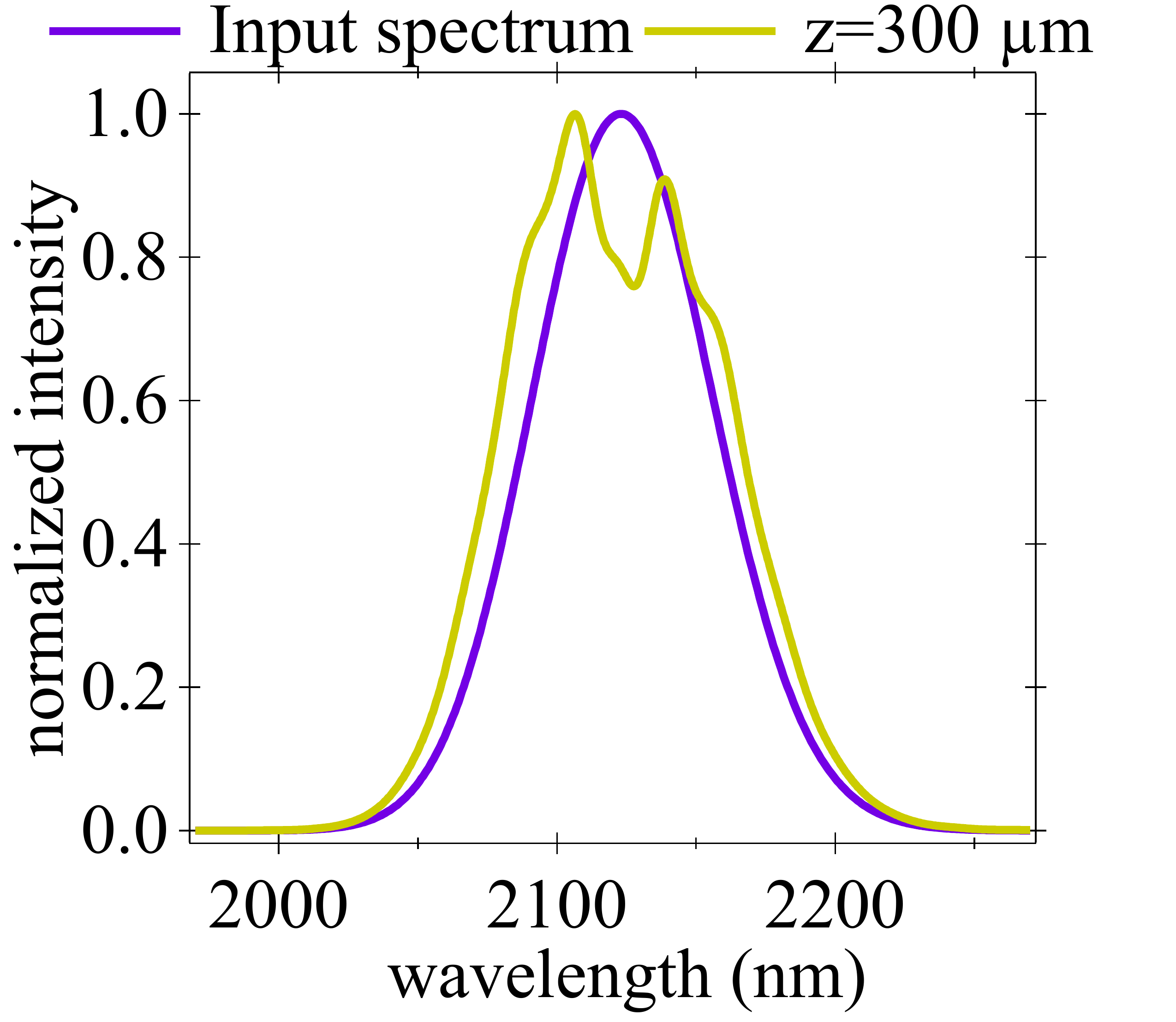}
		    	\caption{}
		    	\label{fig:Si_vs_focus}
		    \end{subfigure}
	        \centering
		    \begin{subfigure}{0.24\textwidth}
			    \includegraphics[width=5.13cm, height=4.2cm]{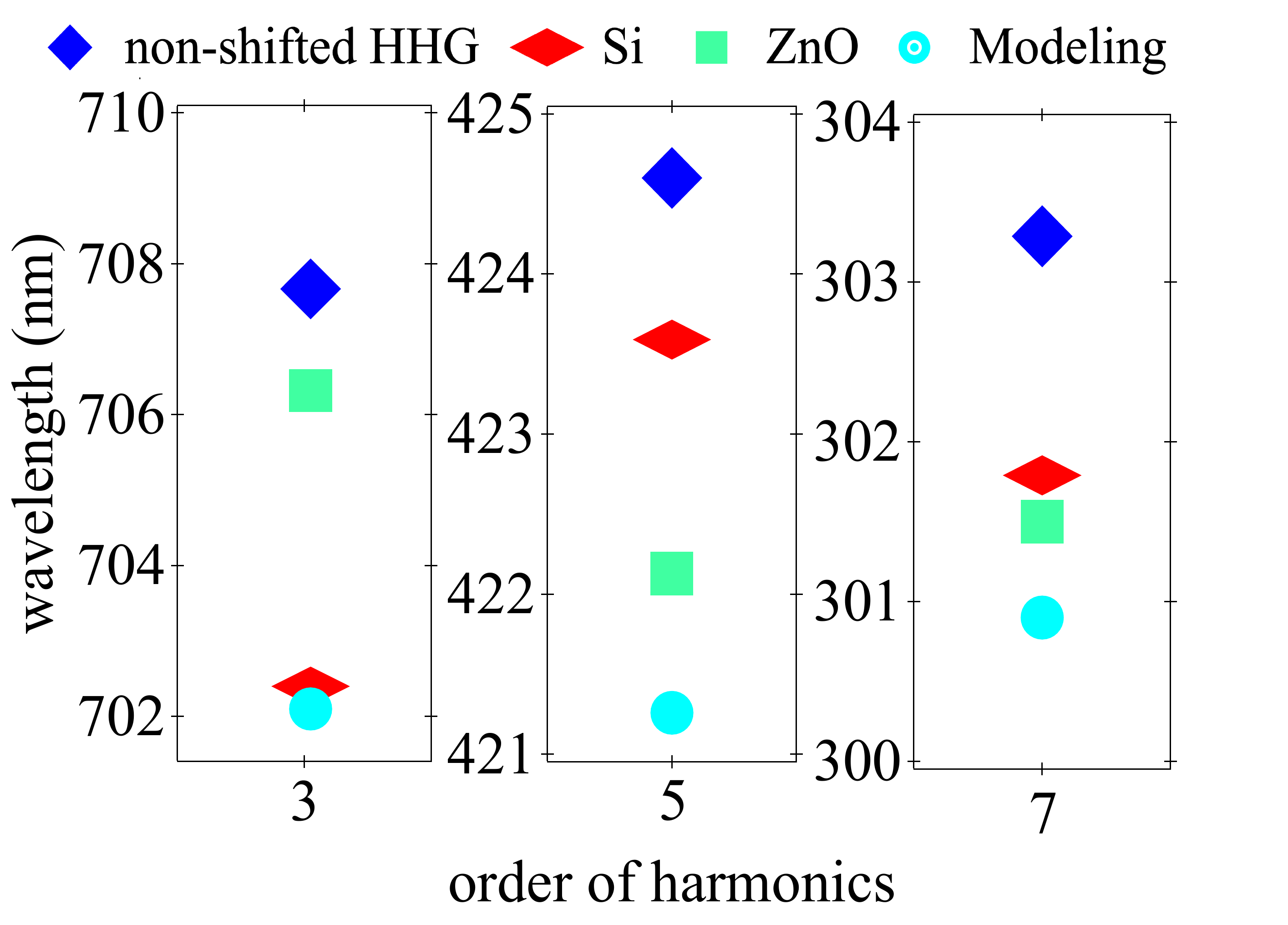}
			    \caption{}
			    \label{fig:Spectrum_Theory_EXP_model}
		    \end{subfigure}
			\caption{(a) Spatially integrated normalized laser spectrum after propagating through Si crystal, input spectrum (purple line) central wavelength at 2.123 $\mathrm{\mu m}$, and for focus position z= 300 $\mathrm{\mu m}$ (yellow line), respectively. (b) Central wavelength dependence of HHG, HHG according to non-shifted driving wavelength (blue cube),  HHG observed in Si (red rhombus), HHG central wavelength observed in ZnO (green square) and as per modelling central wavelength HHG in Si (sky blue circle).}
			    \label{fig:modeling}
\end{figure}

To understand the physical processes that give rise to the harmonic spectral shift, we model the propagation of the driving field inside the Si crystal. We restrict the calculation to z>>z\textsubscript{R}, where z is the focus position inside the crystal and z\textsubscript{R} is the Rayleigh length. We have used the model described in Ref. \cite{couairon2005filamentation}, with the exception that the instantaneous Kerr effect is added to our model. The group velocity dispersion and higher-order dispersion are calculated using a Taylor expansion in  $\omega-\omega\textsubscript{0}$ up to the fifth order of the frequency-dependent $k(\omega$)=$n(\omega)\omega\textsubscript{0}/c$. We then solve the paraxial wave equation in cylindrical geometry, coupled to the evolution equation of electron density for a 85 fs pulse duration, $\sim$ 7 mm Gaussian beam diameter focused onto the 300 $\mathrm{\mu m}$ Si crystal to 2.86 $\mathrm{\mu m}$ by a 3 cm focal length lens. Initially, we have assumed that the conduction band is empty. The spatially integrated laser input spectrum (purple line) and the spectrum obtained after propagation through the 300 $\mathrm{\mu m}$ Si crystal for focus position z=300 $\mathrm{\mu m}$ (yellow line) is shown in Fig. \ref{fig:Si_vs_focus}.

The macroscopic non-linear propagation of the laser pulses in the crystal induces a spectral blueshift of the driving field. This shift will results in a shift of the central wavelength of any high order harmonics generated at the rear side of the crystal. The experimentally measured central wavelengths of the generated harmonics after the propagation of the driving field through a 300 $\mathrm{\mu m} $ thick Si crystal are 702.4 nm and 423.5 nm for H3 and H5, respectively. Fig. \ref{fig:Spectrum_Theory_EXP_model} compare the measured spectral shift at z=0 for the different harmonic orders with the expected values from the numerical propagation of the fundamental pulses. Note that the spectral shift observed for H3 is larger in Si compared to ZnO, while the behaviour is the opposite for higher orders. However, the measured shift for H3 in Si correspond to the expected value, there is a discrepancy between the numerical simulation and the measurement for H5 and H7 as illustrated in Fig. \ref{fig:Spectrum_Theory_EXP_model}.

We attribute the HHG spectral shifts to the non-linear propagation of the intense driving field inside the crystal. Indeed, the strong photoionization contributes to the generation of free-electrons during the transition from the valence band to the conduction band. This induces an intensity-dependent phase delay that results in a blue shift of the generated frequencies. We have found that the distance of propagation inside the crystal is enough to cause a significant blue shift of the HHG spectrum.  This effect is more pronounced when beam is focused at the centre of the crystal. Due to the tight focusing geometry, the intensity dependent interaction region is quite small, therefore the contribution of self-phase modulation would be negligible. Note that it has been reported that interference of H3 generated from the interface and throughout the bulk medium can cause spectral modulations in the third harmonics generated in ZnO \cite{franz2018high}. We have calculated a fringe spacing of 18 nm in 200 $\mathrm{\mu m}$ ZnO while measured peaks are separated by 12 nm. Furthermore, the width of the bandgap is shown to affect the re-absorption of the harmonic spectrum. Wider bandgap materials (such as ZnO used in this study) exhibit less re-absorption and more control of the harmonic spectrum as compared to Si. Through non-linear propagation modelling, we find that the generation of free-carrier density is the principle physical mechanism causing these shifts which induces the change in refractive index due to the contributions of free electron, $n_e(t)\sim$ $1-e^2N_e(t)/(8\pi^2m_e\varepsilon_0 v^2)$, where $m_e$,$e$ are the electron mass and charge, respectively,$\varepsilon_0$ is the vacuum dielectric constant and $N_e(t)$ is the free electron density. During the photoionization process of valence band, $N_e(t)$ increases which result in decreases of $n_e(t)$ and as a result blueshift induces in the harmonics.

\section{Conclusion}
\label{sec:conclusion}

We have observed front and back surface emission due to phase matching of the H3 in ZnO while back surface emission observed due to re-absorption of over bandgap harmonics in Si and ZnO. In ZnO, the generation of the H3 is efficient from the front and rear surfaces of the crystal and a modulated spectrum observed when focusing at the centre of the sample. We attribute this to the interference of harmonics generated at the front surface and in the bulk crystal. We have been demonstrated the changes in the emitted harmonic spectrum, as a function of driving laser focal position in the crystals. The general trend we observe is that the harmonic peaks are blueshifted as the intensity of driving pulses is translated into crystals. Due to the dependence on the focal position, and hence the propagation length in the crystal, we attribute the observed blueshifts to non-linear propagation effects of the driving laser. This is supported by calculations of the non-linear propagation of the driving field in Si which exhibits a blueshift of the central frequency, attributed to the photoionization of valence band electrons promoted to the conduction band by the driving field. Even in samples that strongly absorb the HHG spectrum, we show that since the emitted harmonics come from a small region at the end of the crystal, modulation of the driving pulse prior to the point of generation can be used to control the HHG spectrum. For materials with bandgaps higher than the harmonic photon energy, further control of the spectrum can be obtained from front and back surface emission, as we have demonstrated with the third harmonic in ZnO. This study reveals the fine control of the HHG emission spectrum through macroscopic non-linear effects in the bulk crystal alone. This technique could be used for tailoring frequencies for ultrafast spectroscopic techniques or spectral shaping for attosecond pulse generation.

\section*{Funding}
This work was partially supported by the Fundação para a Ciência e Tecnologia (FCT) under the grant number PD/BD/135224/2017 in the framework of the Advanced Program in Plasma Science and Engineering (APPLAuSE).\\ 
We acknowledge support from the PETACom FET Open H2020 grant number 829153, OPTOLogic FET Open H2020 grant number 899794, from the European Union's Horizon 2020 research and innovation program under grant agreement no 871124 Laserlab-Europe, support from the French ministry of research through the DGA RAPID grant “SWIM”, and the LABEX “PALM” (ANR-10-LABX-0039-PALM) through the grants ”Plasmon-X”, “STAMPS” and “HILAC”. We acknowledge the financial support from the French ASTRE program through the “NanoLight” grant.

%\section*{Acknowledgments}

%\begin{acknowledgments}

%\end{acknowledgments}

%\appendix

%\section{Appendixes}

\nocite{*}
\bibliography{aipsamp}% Produces the bibliography via BibTeX.

\providecommand{\noopsort}[1]{}\providecommand{\singleletter}[1]{#1}%
\begin{thebibliography}{36}%
\makeatletter
\providecommand \@ifxundefined [1]{%
 \@ifx{#1\undefined}
}%
\providecommand \@ifnum [1]{%
 \ifnum #1\expandafter \@firstoftwo
 \else \expandafter \@secondoftwo
 \fi
}%
\providecommand \@ifx [1]{%
 \ifx #1\expandafter \@firstoftwo
 \else \expandafter \@secondoftwo
 \fi
}%
\providecommand \natexlab [1]{#1}%
\providecommand \enquote  [1]{``#1''}%
\providecommand \bibnamefont  [1]{#1}%
\providecommand \bibfnamefont [1]{#1}%
\providecommand \citenamefont [1]{#1}%
\providecommand \href@noop [0]{\@secondoftwo}%
\providecommand \href [0]{\begingroup \@sanitize@url \@href}%
\providecommand \@href[1]{\@@startlink{#1}\@@href}%
\providecommand \@@href[1]{\endgroup#1\@@endlink}%
\providecommand \@sanitize@url [0]{\catcode `\\12\catcode `\$12\catcode
  `\&12\catcode `\#12\catcode `\^12\catcode `\_12\catcode `\%12\relax}%
\providecommand \@@startlink[1]{}%
\providecommand \@@endlink[0]{}%
\providecommand \url  [0]{\begingroup\@sanitize@url \@url }%
\providecommand \@url [1]{\endgroup\@href {#1}{\urlprefix }}%
\providecommand \urlprefix  [0]{URL }%
\providecommand \Eprint [0]{\href }%
\providecommand \doibase [0]{http://dx.doi.org/}%
\providecommand \selectlanguage [0]{\@gobble}%
\providecommand \bibinfo  [0]{\@secondoftwo}%
\providecommand \bibfield  [0]{\@secondoftwo}%
\providecommand \translation [1]{[#1]}%
\providecommand \BibitemOpen [0]{}%
\providecommand \bibitemStop [0]{}%
\providecommand \bibitemNoStop [0]{.\EOS\space}%
\providecommand \EOS [0]{\spacefactor3000\relax}%
\providecommand \BibitemShut  [1]{\csname bibitem#1\endcsname}%
\let\auto@bib@innerbib\@empty
%</preamble>
\bibitem [{\citenamefont {Ghimire}\ \emph {et~al.}(2011)\citenamefont
  {Ghimire}, \citenamefont {DiChiara}, \citenamefont {Sistrunk}, \citenamefont
  {Agostini}, \citenamefont {DiMauro},\ and\ \citenamefont
  {Reis}}]{ghimire2011observation}%
  \BibitemOpen
  \bibfield  {author} {\bibinfo {author} {\bibfnamefont {S.}~\bibnamefont
  {Ghimire}}, \bibinfo {author} {\bibfnamefont {A.~D.}\ \bibnamefont
  {DiChiara}}, \bibinfo {author} {\bibfnamefont {E.}~\bibnamefont {Sistrunk}},
  \bibinfo {author} {\bibfnamefont {P.}~\bibnamefont {Agostini}}, \bibinfo
  {author} {\bibfnamefont {L.~F.}\ \bibnamefont {DiMauro}}, \ and\ \bibinfo
  {author} {\bibfnamefont {D.~A.}\ \bibnamefont {Reis}},\ }\bibfield  {title}
  {\enquote {\bibinfo {title} {Observation of high-order harmonic generation in
  a bulk crystal},}\ }\href@noop {} {\bibfield  {journal} {\bibinfo  {journal}
  {Nature physics}\ }\textbf {\bibinfo {volume} {7}},\ \bibinfo {pages} {138}
  (\bibinfo {year} {2011})}\BibitemShut {NoStop}%
\bibitem [{\citenamefont {You}, \citenamefont {Reis},\ and\ \citenamefont
  {Ghimire}(2017)}]{you2017anisotropic}%
  \BibitemOpen
  \bibfield  {author} {\bibinfo {author} {\bibfnamefont {Y.~S.}\ \bibnamefont
  {You}}, \bibinfo {author} {\bibfnamefont {D.~A.}\ \bibnamefont {Reis}}, \
  and\ \bibinfo {author} {\bibfnamefont {S.}~\bibnamefont {Ghimire}},\
  }\bibfield  {title} {\enquote {\bibinfo {title} {Anisotropic high-harmonic
  generation in bulk crystals},}\ }\href@noop {} {\bibfield  {journal}
  {\bibinfo  {journal} {Nature Physics}\ }\textbf {\bibinfo {volume} {13}},\
  \bibinfo {pages} {345} (\bibinfo {year} {2017})}\BibitemShut {NoStop}%
\bibitem [{\citenamefont {You}\ \emph {et~al.}(2017)\citenamefont {You},
  \citenamefont {Yin}, \citenamefont {Wu}, \citenamefont {Chew}, \citenamefont
  {Ren}, \citenamefont {Zhuang}, \citenamefont {Gholam-Mirzaei}, \citenamefont
  {Chini}, \citenamefont {Chang},\ and\ \citenamefont {Ghimire}}]{you2017high}%
  \BibitemOpen
  \bibfield  {author} {\bibinfo {author} {\bibfnamefont {Y.~S.}\ \bibnamefont
  {You}}, \bibinfo {author} {\bibfnamefont {Y.}~\bibnamefont {Yin}}, \bibinfo
  {author} {\bibfnamefont {Y.}~\bibnamefont {Wu}}, \bibinfo {author}
  {\bibfnamefont {A.}~\bibnamefont {Chew}}, \bibinfo {author} {\bibfnamefont
  {X.}~\bibnamefont {Ren}}, \bibinfo {author} {\bibfnamefont {F.}~\bibnamefont
  {Zhuang}}, \bibinfo {author} {\bibfnamefont {S.}~\bibnamefont
  {Gholam-Mirzaei}}, \bibinfo {author} {\bibfnamefont {M.}~\bibnamefont
  {Chini}}, \bibinfo {author} {\bibfnamefont {Z.}~\bibnamefont {Chang}}, \ and\
  \bibinfo {author} {\bibfnamefont {S.}~\bibnamefont {Ghimire}},\ }\bibfield
  {title} {\enquote {\bibinfo {title} {High-harmonic generation in amorphous
  solids},}\ }\href@noop {} {\bibfield  {journal} {\bibinfo  {journal} {Nature
  communications}\ }\textbf {\bibinfo {volume} {8}},\ \bibinfo {pages} {724}
  (\bibinfo {year} {2017})}\BibitemShut {NoStop}%
\bibitem [{\citenamefont {Golde}, \citenamefont {Meier},\ and\ \citenamefont
  {Koch}(2008)}]{golde2008high}%
  \BibitemOpen
  \bibfield  {author} {\bibinfo {author} {\bibfnamefont {D.}~\bibnamefont
  {Golde}}, \bibinfo {author} {\bibfnamefont {T.}~\bibnamefont {Meier}}, \ and\
  \bibinfo {author} {\bibfnamefont {S.~W.}\ \bibnamefont {Koch}},\ }\bibfield
  {title} {\enquote {\bibinfo {title} {High harmonics generated in
  semiconductor nanostructures by the coupled dynamics of optical inter-and
  intraband excitations},}\ }\href@noop {} {\bibfield  {journal} {\bibinfo
  {journal} {Physical Review B}\ }\textbf {\bibinfo {volume} {77}},\ \bibinfo
  {pages} {075330} (\bibinfo {year} {2008})}\BibitemShut {NoStop}%
\bibitem [{\citenamefont {Higuchi}, \citenamefont {Stockman},\ and\
  \citenamefont {Hommelhoff}(2014)}]{higuchi2014strong}%
  \BibitemOpen
  \bibfield  {author} {\bibinfo {author} {\bibfnamefont {T.}~\bibnamefont
  {Higuchi}}, \bibinfo {author} {\bibfnamefont {M.~I.}\ \bibnamefont
  {Stockman}}, \ and\ \bibinfo {author} {\bibfnamefont {P.}~\bibnamefont
  {Hommelhoff}},\ }\bibfield  {title} {\enquote {\bibinfo {title} {Strong-field
  perspective on high-harmonic radiation from bulk solids},}\ }\href@noop {}
  {\bibfield  {journal} {\bibinfo  {journal} {Physical review letters}\
  }\textbf {\bibinfo {volume} {113}},\ \bibinfo {pages} {213901} (\bibinfo
  {year} {2014})}\BibitemShut {NoStop}%
\bibitem [{\citenamefont {Hawkins}, \citenamefont {Ivanov},\ and\ \citenamefont
  {Yakovlev}(2015)}]{hawkins2015effect}%
  \BibitemOpen
  \bibfield  {author} {\bibinfo {author} {\bibfnamefont {P.~G.}\ \bibnamefont
  {Hawkins}}, \bibinfo {author} {\bibfnamefont {M.~Y.}\ \bibnamefont {Ivanov}},
  \ and\ \bibinfo {author} {\bibfnamefont {V.~S.}\ \bibnamefont {Yakovlev}},\
  }\bibfield  {title} {\enquote {\bibinfo {title} {Effect of multiple
  conduction bands on high-harmonic emission from dielectrics},}\ }\href@noop
  {} {\bibfield  {journal} {\bibinfo  {journal} {Physical Review A}\ }\textbf
  {\bibinfo {volume} {91}},\ \bibinfo {pages} {013405} (\bibinfo {year}
  {2015})}\BibitemShut {NoStop}%
\bibitem [{\citenamefont {Wu}\ \emph {et~al.}(2016)\citenamefont {Wu},
  \citenamefont {Browne}, \citenamefont {Schafer},\ and\ \citenamefont
  {Gaarde}}]{wu2016multilevel}%
  \BibitemOpen
  \bibfield  {author} {\bibinfo {author} {\bibfnamefont {M.}~\bibnamefont
  {Wu}}, \bibinfo {author} {\bibfnamefont {D.~A.}\ \bibnamefont {Browne}},
  \bibinfo {author} {\bibfnamefont {K.~J.}\ \bibnamefont {Schafer}}, \ and\
  \bibinfo {author} {\bibfnamefont {M.~B.}\ \bibnamefont {Gaarde}},\ }\bibfield
   {title} {\enquote {\bibinfo {title} {Multilevel perspective on high-order
  harmonic generation in solids},}\ }\href@noop {} {\bibfield  {journal}
  {\bibinfo  {journal} {Physical Review A}\ }\textbf {\bibinfo {volume} {94}},\
  \bibinfo {pages} {063403} (\bibinfo {year} {2016})}\BibitemShut {NoStop}%
\bibitem [{\citenamefont {Tancogne-Dejean}\ \emph {et~al.}(2017)\citenamefont
  {Tancogne-Dejean}, \citenamefont {M{\"u}cke}, \citenamefont {K{\"a}rtner},\
  and\ \citenamefont {Rubio}}]{tancogne2017impact}%
  \BibitemOpen
  \bibfield  {author} {\bibinfo {author} {\bibfnamefont {N.}~\bibnamefont
  {Tancogne-Dejean}}, \bibinfo {author} {\bibfnamefont {O.~D.}\ \bibnamefont
  {M{\"u}cke}}, \bibinfo {author} {\bibfnamefont {F.~X.}\ \bibnamefont
  {K{\"a}rtner}}, \ and\ \bibinfo {author} {\bibfnamefont {A.}~\bibnamefont
  {Rubio}},\ }\bibfield  {title} {\enquote {\bibinfo {title} {Impact of the
  electronic band structure in high-harmonic generation spectra of solids},}\
  }\href@noop {} {\bibfield  {journal} {\bibinfo  {journal} {Physical review
  letters}\ }\textbf {\bibinfo {volume} {118}},\ \bibinfo {pages} {087403}
  (\bibinfo {year} {2017})}\BibitemShut {NoStop}%
\bibitem [{\citenamefont {Floss}\ \emph {et~al.}(2018)\citenamefont {Floss},
  \citenamefont {Lemell}, \citenamefont {Wachter}, \citenamefont {Smejkal},
  \citenamefont {Sato}, \citenamefont {Tong}, \citenamefont {Yabana},\ and\
  \citenamefont {Burgd{\"o}rfer}}]{floss2018ab}%
  \BibitemOpen
  \bibfield  {author} {\bibinfo {author} {\bibfnamefont {I.}~\bibnamefont
  {Floss}}, \bibinfo {author} {\bibfnamefont {C.}~\bibnamefont {Lemell}},
  \bibinfo {author} {\bibfnamefont {G.}~\bibnamefont {Wachter}}, \bibinfo
  {author} {\bibfnamefont {V.}~\bibnamefont {Smejkal}}, \bibinfo {author}
  {\bibfnamefont {S.~A.}\ \bibnamefont {Sato}}, \bibinfo {author}
  {\bibfnamefont {X.-M.}\ \bibnamefont {Tong}}, \bibinfo {author}
  {\bibfnamefont {K.}~\bibnamefont {Yabana}}, \ and\ \bibinfo {author}
  {\bibfnamefont {J.}~\bibnamefont {Burgd{\"o}rfer}},\ }\bibfield  {title}
  {\enquote {\bibinfo {title} {Ab initio multiscale simulation of high-order
  harmonic generation in solids},}\ }\href@noop {} {\bibfield  {journal}
  {\bibinfo  {journal} {Physical Review A}\ }\textbf {\bibinfo {volume} {97}},\
  \bibinfo {pages} {011401} (\bibinfo {year} {2018})}\BibitemShut {NoStop}%
\bibitem [{\citenamefont {Marangos}(2011)}]{marangos2011high}%
  \BibitemOpen
  \bibfield  {author} {\bibinfo {author} {\bibfnamefont {J.~P.}\ \bibnamefont
  {Marangos}},\ }\bibfield  {title} {\enquote {\bibinfo {title} {High-harmonic
  generation: Solid progress},}\ }\href@noop {} {\bibfield  {journal} {\bibinfo
   {journal} {Nature Physics}\ }\textbf {\bibinfo {volume} {7}},\ \bibinfo
  {pages} {97} (\bibinfo {year} {2011})}\BibitemShut {NoStop}%
\bibitem [{\citenamefont {Schubert}\ \emph {et~al.}(2014)\citenamefont
  {Schubert}, \citenamefont {Hohenleutner}, \citenamefont {Langer},
  \citenamefont {Urbanek}, \citenamefont {Lange}, \citenamefont {Huttner},
  \citenamefont {Golde}, \citenamefont {Meier}, \citenamefont {Kira},
  \citenamefont {Koch} \emph {et~al.}}]{schubert2014sub}%
  \BibitemOpen
  \bibfield  {author} {\bibinfo {author} {\bibfnamefont {O.}~\bibnamefont
  {Schubert}}, \bibinfo {author} {\bibfnamefont {M.}~\bibnamefont
  {Hohenleutner}}, \bibinfo {author} {\bibfnamefont {F.}~\bibnamefont
  {Langer}}, \bibinfo {author} {\bibfnamefont {B.}~\bibnamefont {Urbanek}},
  \bibinfo {author} {\bibfnamefont {C.}~\bibnamefont {Lange}}, \bibinfo
  {author} {\bibfnamefont {U.}~\bibnamefont {Huttner}}, \bibinfo {author}
  {\bibfnamefont {D.}~\bibnamefont {Golde}}, \bibinfo {author} {\bibfnamefont
  {T.}~\bibnamefont {Meier}}, \bibinfo {author} {\bibfnamefont
  {M.}~\bibnamefont {Kira}}, \bibinfo {author} {\bibfnamefont {S.~W.}\
  \bibnamefont {Koch}},  \emph {et~al.},\ }\bibfield  {title} {\enquote
  {\bibinfo {title} {Sub-cycle control of terahertz high-harmonic generation by
  dynamical bloch oscillations},}\ }\href@noop {} {\bibfield  {journal}
  {\bibinfo  {journal} {Nature Photonics}\ }\textbf {\bibinfo {volume} {8}},\
  \bibinfo {pages} {119} (\bibinfo {year} {2014})}\BibitemShut {NoStop}%
\bibitem [{\citenamefont {Hohenleutner}\ \emph {et~al.}(2015)\citenamefont
  {Hohenleutner}, \citenamefont {Langer}, \citenamefont {Schubert},
  \citenamefont {Knorr}, \citenamefont {Huttner}, \citenamefont {Koch},
  \citenamefont {Kira},\ and\ \citenamefont {Huber}}]{hohenleutner2015real}%
  \BibitemOpen
  \bibfield  {author} {\bibinfo {author} {\bibfnamefont {M.}~\bibnamefont
  {Hohenleutner}}, \bibinfo {author} {\bibfnamefont {F.}~\bibnamefont
  {Langer}}, \bibinfo {author} {\bibfnamefont {O.}~\bibnamefont {Schubert}},
  \bibinfo {author} {\bibfnamefont {M.}~\bibnamefont {Knorr}}, \bibinfo
  {author} {\bibfnamefont {U.}~\bibnamefont {Huttner}}, \bibinfo {author}
  {\bibfnamefont {S.}~\bibnamefont {Koch}}, \bibinfo {author} {\bibfnamefont
  {M.}~\bibnamefont {Kira}}, \ and\ \bibinfo {author} {\bibfnamefont
  {R.}~\bibnamefont {Huber}},\ }\bibfield  {title} {\enquote {\bibinfo {title}
  {Real-time observation of interfering crystal electrons in high-harmonic
  generation},}\ }\href@noop {} {\bibfield  {journal} {\bibinfo  {journal}
  {Nature}\ }\textbf {\bibinfo {volume} {523}},\ \bibinfo {pages} {572}
  (\bibinfo {year} {2015})}\BibitemShut {NoStop}%
\bibitem [{\citenamefont {Luu}\ \emph {et~al.}(2015)\citenamefont {Luu},
  \citenamefont {Garg}, \citenamefont {Kruchinin}, \citenamefont {Moulet},
  \citenamefont {Hassan},\ and\ \citenamefont {Goulielmakis}}]{luu2015extreme}%
  \BibitemOpen
  \bibfield  {author} {\bibinfo {author} {\bibfnamefont {T.~T.}\ \bibnamefont
  {Luu}}, \bibinfo {author} {\bibfnamefont {M.}~\bibnamefont {Garg}}, \bibinfo
  {author} {\bibfnamefont {S.~Y.}\ \bibnamefont {Kruchinin}}, \bibinfo {author}
  {\bibfnamefont {A.}~\bibnamefont {Moulet}}, \bibinfo {author} {\bibfnamefont
  {M.~T.}\ \bibnamefont {Hassan}}, \ and\ \bibinfo {author} {\bibfnamefont
  {E.}~\bibnamefont {Goulielmakis}},\ }\bibfield  {title} {\enquote {\bibinfo
  {title} {Extreme ultraviolet high-harmonic spectroscopy of solids},}\
  }\href@noop {} {\bibfield  {journal} {\bibinfo  {journal} {Nature}\ }\textbf
  {\bibinfo {volume} {521}},\ \bibinfo {pages} {498} (\bibinfo {year}
  {2015})}\BibitemShut {NoStop}%
\bibitem [{\citenamefont {Vampa}\ \emph
  {et~al.}(2015{\natexlab{a}})\citenamefont {Vampa}, \citenamefont {Hammond},
  \citenamefont {Thir{\'e}}, \citenamefont {Schmidt}, \citenamefont
  {L{\'e}gar{\'e}}, \citenamefont {McDonald}, \citenamefont {Brabec},
  \citenamefont {Klug},\ and\ \citenamefont {Corkum}}]{vampa2015all}%
  \BibitemOpen
  \bibfield  {author} {\bibinfo {author} {\bibfnamefont {G.}~\bibnamefont
  {Vampa}}, \bibinfo {author} {\bibfnamefont {T.}~\bibnamefont {Hammond}},
  \bibinfo {author} {\bibfnamefont {N.}~\bibnamefont {Thir{\'e}}}, \bibinfo
  {author} {\bibfnamefont {B.}~\bibnamefont {Schmidt}}, \bibinfo {author}
  {\bibfnamefont {F.}~\bibnamefont {L{\'e}gar{\'e}}}, \bibinfo {author}
  {\bibfnamefont {C.}~\bibnamefont {McDonald}}, \bibinfo {author}
  {\bibfnamefont {T.}~\bibnamefont {Brabec}}, \bibinfo {author} {\bibfnamefont
  {D.}~\bibnamefont {Klug}}, \ and\ \bibinfo {author} {\bibfnamefont
  {P.}~\bibnamefont {Corkum}},\ }\bibfield  {title} {\enquote {\bibinfo {title}
  {All-optical reconstruction of crystal band structure},}\ }\href@noop {}
  {\bibfield  {journal} {\bibinfo  {journal} {Physical review letters}\
  }\textbf {\bibinfo {volume} {115}},\ \bibinfo {pages} {193603} (\bibinfo
  {year} {2015}{\natexlab{a}})}\BibitemShut {NoStop}%
\bibitem [{\citenamefont {Ghimire}\ and\ \citenamefont
  {Reis}(2018)}]{ghimire2018high}%
  \BibitemOpen
  \bibfield  {author} {\bibinfo {author} {\bibfnamefont {S.}~\bibnamefont
  {Ghimire}}\ and\ \bibinfo {author} {\bibfnamefont {D.~A.}\ \bibnamefont
  {Reis}},\ }\bibfield  {title} {\enquote {\bibinfo {title} {High-harmonic
  generation from solids},}\ }\href@noop {} {\bibfield  {journal} {\bibinfo
  {journal} {Nature Physics}\ ,\ \bibinfo {pages} {1}} (\bibinfo {year}
  {2018})}\BibitemShut {NoStop}%
\bibitem [{\citenamefont {Vampa}\ and\ \citenamefont
  {Brabec}(2017)}]{vampa2017merge}%
  \BibitemOpen
  \bibfield  {author} {\bibinfo {author} {\bibfnamefont {G.}~\bibnamefont
  {Vampa}}\ and\ \bibinfo {author} {\bibfnamefont {T.}~\bibnamefont {Brabec}},\
  }\bibfield  {title} {\enquote {\bibinfo {title} {Merge of high harmonic
  generation from gases and solids and its implications for attosecond
  science},}\ }\href@noop {} {\bibfield  {journal} {\bibinfo  {journal}
  {Journal of Physics B: Atomic, Molecular and Optical Physics}\ }\textbf
  {\bibinfo {volume} {50}},\ \bibinfo {pages} {083001} (\bibinfo {year}
  {2017})}\BibitemShut {NoStop}%
\bibitem [{\citenamefont {Vampa}\ \emph
  {et~al.}(2015{\natexlab{b}})\citenamefont {Vampa}, \citenamefont {McDonald},
  \citenamefont {Orlando}, \citenamefont {Corkum},\ and\ \citenamefont
  {Brabec}}]{vampa2015semiclassical}%
  \BibitemOpen
  \bibfield  {author} {\bibinfo {author} {\bibfnamefont {G.}~\bibnamefont
  {Vampa}}, \bibinfo {author} {\bibfnamefont {C.}~\bibnamefont {McDonald}},
  \bibinfo {author} {\bibfnamefont {G.}~\bibnamefont {Orlando}}, \bibinfo
  {author} {\bibfnamefont {P.}~\bibnamefont {Corkum}}, \ and\ \bibinfo {author}
  {\bibfnamefont {T.}~\bibnamefont {Brabec}},\ }\bibfield  {title} {\enquote
  {\bibinfo {title} {Semiclassical analysis of high harmonic generation in bulk
  crystals},}\ }\href@noop {} {\bibfield  {journal} {\bibinfo  {journal}
  {Physical Review B}\ }\textbf {\bibinfo {volume} {91}},\ \bibinfo {pages}
  {064302} (\bibinfo {year} {2015}{\natexlab{b}})}\BibitemShut {NoStop}%
\bibitem [{\citenamefont {Gholam-Mirzaei}, \citenamefont {Beetar},\ and\
  \citenamefont {Chini}(2017)}]{gholam2017high}%
  \BibitemOpen
  \bibfield  {author} {\bibinfo {author} {\bibfnamefont {S.}~\bibnamefont
  {Gholam-Mirzaei}}, \bibinfo {author} {\bibfnamefont {J.}~\bibnamefont
  {Beetar}}, \ and\ \bibinfo {author} {\bibfnamefont {M.}~\bibnamefont
  {Chini}},\ }\bibfield  {title} {\enquote {\bibinfo {title} {High harmonic
  generation in zno with a high-power mid-ir opa},}\ }\href@noop {} {\bibfield
  {journal} {\bibinfo  {journal} {Applied Physics Letters}\ }\textbf {\bibinfo
  {volume} {110}},\ \bibinfo {pages} {061101} (\bibinfo {year}
  {2017})}\BibitemShut {NoStop}%
\bibitem [{\citenamefont {Gholam-Mirzaei}\ \emph {et~al.}(2018)\citenamefont
  {Gholam-Mirzaei}, \citenamefont {Beetar}, \citenamefont {Chac{\'o}n},\ and\
  \citenamefont {Chini}}]{gholam2018high}%
  \BibitemOpen
  \bibfield  {author} {\bibinfo {author} {\bibfnamefont {S.}~\bibnamefont
  {Gholam-Mirzaei}}, \bibinfo {author} {\bibfnamefont {J.~E.}\ \bibnamefont
  {Beetar}}, \bibinfo {author} {\bibfnamefont {A.}~\bibnamefont {Chac{\'o}n}},
  \ and\ \bibinfo {author} {\bibfnamefont {M.}~\bibnamefont {Chini}},\
  }\bibfield  {title} {\enquote {\bibinfo {title} {High-harmonic generation in
  zno driven by self-compressed mid-infrared pulses},}\ }\href@noop {}
  {\bibfield  {journal} {\bibinfo  {journal} {JOSA B}\ }\textbf {\bibinfo
  {volume} {35}},\ \bibinfo {pages} {A27--A31} (\bibinfo {year}
  {2018})}\BibitemShut {NoStop}%
\bibitem [{\citenamefont {Vampa}\ \emph {et~al.}(2016)\citenamefont {Vampa},
  \citenamefont {Ghamsari}, \citenamefont {Mousavi}, \citenamefont {Hammond},
  \citenamefont {Olivieri}, \citenamefont {Lisicka-Skrek}, \citenamefont
  {Naumov}, \citenamefont {Villeneuve}, \citenamefont {Staudte}, \citenamefont
  {Berini} \emph {et~al.}}]{vampa2016plasmonic}%
  \BibitemOpen
  \bibfield  {author} {\bibinfo {author} {\bibfnamefont {G.}~\bibnamefont
  {Vampa}}, \bibinfo {author} {\bibfnamefont {B.}~\bibnamefont {Ghamsari}},
  \bibinfo {author} {\bibfnamefont {S.~S.}\ \bibnamefont {Mousavi}}, \bibinfo
  {author} {\bibfnamefont {T.}~\bibnamefont {Hammond}}, \bibinfo {author}
  {\bibfnamefont {A.}~\bibnamefont {Olivieri}}, \bibinfo {author}
  {\bibfnamefont {E.}~\bibnamefont {Lisicka-Skrek}}, \bibinfo {author}
  {\bibfnamefont {A.}~\bibnamefont {Naumov}}, \bibinfo {author} {\bibfnamefont
  {D.}~\bibnamefont {Villeneuve}}, \bibinfo {author} {\bibfnamefont
  {A.}~\bibnamefont {Staudte}}, \bibinfo {author} {\bibfnamefont
  {P.}~\bibnamefont {Berini}},  \emph {et~al.},\ }\bibfield  {title} {\enquote
  {\bibinfo {title} {Plasmonic-enhanced high harmonic generation from bulk
  silicon},}\ }in\ \href@noop {} {\emph {\bibinfo {booktitle} {Laser
  Science}}}\ (\bibinfo {organization} {Optical Society of America},\ \bibinfo
  {year} {2016})\ pp.\ \bibinfo {pages} {LTh5I--2}\BibitemShut {NoStop}%
\bibitem [{\citenamefont {Vampa}\ \emph {et~al.}(2017)\citenamefont {Vampa},
  \citenamefont {Ghamsari}, \citenamefont {Mousavi}, \citenamefont {Hammond},
  \citenamefont {Olivieri}, \citenamefont {Lisicka-Skrek}, \citenamefont
  {Naumov}, \citenamefont {Villeneuve}, \citenamefont {Staudte}, \citenamefont
  {Berini} \emph {et~al.}}]{vampa2017plasmon}%
  \BibitemOpen
  \bibfield  {author} {\bibinfo {author} {\bibfnamefont {G.}~\bibnamefont
  {Vampa}}, \bibinfo {author} {\bibfnamefont {B.}~\bibnamefont {Ghamsari}},
  \bibinfo {author} {\bibfnamefont {S.~S.}\ \bibnamefont {Mousavi}}, \bibinfo
  {author} {\bibfnamefont {T.}~\bibnamefont {Hammond}}, \bibinfo {author}
  {\bibfnamefont {A.}~\bibnamefont {Olivieri}}, \bibinfo {author}
  {\bibfnamefont {E.}~\bibnamefont {Lisicka-Skrek}}, \bibinfo {author}
  {\bibfnamefont {A.~Y.}\ \bibnamefont {Naumov}}, \bibinfo {author}
  {\bibfnamefont {D.}~\bibnamefont {Villeneuve}}, \bibinfo {author}
  {\bibfnamefont {A.}~\bibnamefont {Staudte}}, \bibinfo {author} {\bibfnamefont
  {P.}~\bibnamefont {Berini}},  \emph {et~al.},\ }\bibfield  {title} {\enquote
  {\bibinfo {title} {Plasmon-enhanced high-harmonic generation from silicon},}\
  }\href@noop {} {\bibfield  {journal} {\bibinfo  {journal} {Nature Physics}\
  }\textbf {\bibinfo {volume} {13}},\ \bibinfo {pages} {659--662} (\bibinfo
  {year} {2017})}\BibitemShut {NoStop}%
\bibitem [{\citenamefont {Vampa}\ \emph {et~al.}(2019)\citenamefont {Vampa},
  \citenamefont {Liu}, \citenamefont {Heinz},\ and\ \citenamefont
  {Reis}}]{vampa2019disentangling}%
  \BibitemOpen
  \bibfield  {author} {\bibinfo {author} {\bibfnamefont {G.}~\bibnamefont
  {Vampa}}, \bibinfo {author} {\bibfnamefont {H.}~\bibnamefont {Liu}}, \bibinfo
  {author} {\bibfnamefont {T.~F.}\ \bibnamefont {Heinz}}, \ and\ \bibinfo
  {author} {\bibfnamefont {D.~A.}\ \bibnamefont {Reis}},\ }\bibfield  {title}
  {\enquote {\bibinfo {title} {Disentangling interface and bulk contributions
  to high-harmonic emission from solids},}\ }\href@noop {} {\bibfield
  {journal} {\bibinfo  {journal} {Optica}\ }\textbf {\bibinfo {volume} {6}},\
  \bibinfo {pages} {553--556} (\bibinfo {year} {2019})}\BibitemShut {NoStop}%
\bibitem [{\citenamefont {Sivis}\ \emph {et~al.}(2017)\citenamefont {Sivis},
  \citenamefont {Taucer}, \citenamefont {Vampa}, \citenamefont {Johnston},
  \citenamefont {Staudte}, \citenamefont {Naumov}, \citenamefont {Villeneuve},
  \citenamefont {Ropers},\ and\ \citenamefont {Corkum}}]{sivis2017tailored}%
  \BibitemOpen
  \bibfield  {author} {\bibinfo {author} {\bibfnamefont {M.}~\bibnamefont
  {Sivis}}, \bibinfo {author} {\bibfnamefont {M.}~\bibnamefont {Taucer}},
  \bibinfo {author} {\bibfnamefont {G.}~\bibnamefont {Vampa}}, \bibinfo
  {author} {\bibfnamefont {K.}~\bibnamefont {Johnston}}, \bibinfo {author}
  {\bibfnamefont {A.}~\bibnamefont {Staudte}}, \bibinfo {author} {\bibfnamefont
  {A.~Y.}\ \bibnamefont {Naumov}}, \bibinfo {author} {\bibfnamefont
  {D.}~\bibnamefont {Villeneuve}}, \bibinfo {author} {\bibfnamefont
  {C.}~\bibnamefont {Ropers}}, \ and\ \bibinfo {author} {\bibfnamefont
  {P.}~\bibnamefont {Corkum}},\ }\bibfield  {title} {\enquote {\bibinfo {title}
  {Tailored semiconductors for high-harmonic optoelectronics},}\ }\href@noop {}
  {\bibfield  {journal} {\bibinfo  {journal} {Science}\ }\textbf {\bibinfo
  {volume} {357}},\ \bibinfo {pages} {303--306} (\bibinfo {year}
  {2017})}\BibitemShut {NoStop}%
\bibitem [{\citenamefont {Franz}\ \emph {et~al.}(2019)\citenamefont {Franz},
  \citenamefont {Kaassamani}, \citenamefont {Gauthier}, \citenamefont
  {Nicolas}, \citenamefont {Kholodtsova}, \citenamefont {Douillard},
  \citenamefont {Gomes}, \citenamefont {Lavoute}, \citenamefont {Gaponov},
  \citenamefont {Ducros} \emph {et~al.}}]{franz2019all}%
  \BibitemOpen
  \bibfield  {author} {\bibinfo {author} {\bibfnamefont {D.}~\bibnamefont
  {Franz}}, \bibinfo {author} {\bibfnamefont {S.}~\bibnamefont {Kaassamani}},
  \bibinfo {author} {\bibfnamefont {D.}~\bibnamefont {Gauthier}}, \bibinfo
  {author} {\bibfnamefont {R.}~\bibnamefont {Nicolas}}, \bibinfo {author}
  {\bibfnamefont {M.}~\bibnamefont {Kholodtsova}}, \bibinfo {author}
  {\bibfnamefont {L.}~\bibnamefont {Douillard}}, \bibinfo {author}
  {\bibfnamefont {J.-T.}\ \bibnamefont {Gomes}}, \bibinfo {author}
  {\bibfnamefont {L.}~\bibnamefont {Lavoute}}, \bibinfo {author} {\bibfnamefont
  {D.}~\bibnamefont {Gaponov}}, \bibinfo {author} {\bibfnamefont
  {N.}~\bibnamefont {Ducros}},  \emph {et~al.},\ }\bibfield  {title} {\enquote
  {\bibinfo {title} {All semiconductor enhanced high-harmonic generation from a
  single nanostructured cone},}\ }\href@noop {} {\bibfield  {journal} {\bibinfo
   {journal} {Scientific reports}\ }\textbf {\bibinfo {volume} {9}},\ \bibinfo
  {pages} {5663} (\bibinfo {year} {2019})}\BibitemShut {NoStop}%
\bibitem [{\citenamefont {Gauthier}\ \emph {et~al.}(2019)\citenamefont
  {Gauthier}, \citenamefont {Kaassamani}, \citenamefont {Franz}, \citenamefont
  {Nicolas}, \citenamefont {Gomes}, \citenamefont {Lavoute}, \citenamefont
  {Gaponov}, \citenamefont {F{\'e}vrier}, \citenamefont {Jargot}, \citenamefont
  {Hanna} \emph {et~al.}}]{gauthier2019orbital}%
  \BibitemOpen
  \bibfield  {author} {\bibinfo {author} {\bibfnamefont {D.}~\bibnamefont
  {Gauthier}}, \bibinfo {author} {\bibfnamefont {S.}~\bibnamefont
  {Kaassamani}}, \bibinfo {author} {\bibfnamefont {D.}~\bibnamefont {Franz}},
  \bibinfo {author} {\bibfnamefont {R.}~\bibnamefont {Nicolas}}, \bibinfo
  {author} {\bibfnamefont {J.-T.}\ \bibnamefont {Gomes}}, \bibinfo {author}
  {\bibfnamefont {L.}~\bibnamefont {Lavoute}}, \bibinfo {author} {\bibfnamefont
  {D.}~\bibnamefont {Gaponov}}, \bibinfo {author} {\bibfnamefont
  {S.}~\bibnamefont {F{\'e}vrier}}, \bibinfo {author} {\bibfnamefont
  {G.}~\bibnamefont {Jargot}}, \bibinfo {author} {\bibfnamefont
  {M.}~\bibnamefont {Hanna}},  \emph {et~al.},\ }\bibfield  {title} {\enquote
  {\bibinfo {title} {Orbital angular momentum from semiconductor high-order
  harmonics},}\ }\href@noop {} {\bibfield  {journal} {\bibinfo  {journal}
  {Optics letters}\ }\textbf {\bibinfo {volume} {44}},\ \bibinfo {pages}
  {546--549} (\bibinfo {year} {2019})}\BibitemShut {NoStop}%
\bibitem [{\citenamefont {Torres}\ and\ \citenamefont
  {Torner}(2011)}]{torres2011twisted}%
  \BibitemOpen
  \bibfield  {author} {\bibinfo {author} {\bibfnamefont {J.~P.}\ \bibnamefont
  {Torres}}\ and\ \bibinfo {author} {\bibfnamefont {L.}~\bibnamefont
  {Torner}},\ }\href@noop {} {\emph {\bibinfo {title} {Twisted photons:
  applications of light with orbital angular momentum}}}\ (\bibinfo
  {publisher} {John Wiley \& Sons},\ \bibinfo {year} {2011})\BibitemShut
  {NoStop}%
\bibitem [{\citenamefont {Brandi}, \citenamefont {Giammanco},\ and\
  \citenamefont {Ubachs}(2006)}]{brandi2006spectral}%
  \BibitemOpen
  \bibfield  {author} {\bibinfo {author} {\bibfnamefont {F.}~\bibnamefont
  {Brandi}}, \bibinfo {author} {\bibfnamefont {F.}~\bibnamefont {Giammanco}}, \
  and\ \bibinfo {author} {\bibfnamefont {W.}~\bibnamefont {Ubachs}},\
  }\bibfield  {title} {\enquote {\bibinfo {title} {Spectral redshift in
  harmonic generation from plasma dynamics in the laser focus},}\ }\href@noop
  {} {\bibfield  {journal} {\bibinfo  {journal} {Physical review letters}\
  }\textbf {\bibinfo {volume} {96}},\ \bibinfo {pages} {123904} (\bibinfo
  {year} {2006})}\BibitemShut {NoStop}%
\bibitem [{\citenamefont {Bian}\ and\ \citenamefont
  {Bandrauk}(2013)}]{bian2013spectral}%
  \BibitemOpen
  \bibfield  {author} {\bibinfo {author} {\bibfnamefont {X.-B.}\ \bibnamefont
  {Bian}}\ and\ \bibinfo {author} {\bibfnamefont {A.~D.}\ \bibnamefont
  {Bandrauk}},\ }\bibfield  {title} {\enquote {\bibinfo {title} {Spectral
  shifts of nonadiabatic high-order harmonic generation},}\ }\href@noop {}
  {\bibfield  {journal} {\bibinfo  {journal} {Applied Sciences}\ }\textbf
  {\bibinfo {volume} {3}},\ \bibinfo {pages} {267--277} (\bibinfo {year}
  {2013})}\BibitemShut {NoStop}%
\bibitem [{\citenamefont {Du}\ \emph {et~al.}(2015)\citenamefont {Du},
  \citenamefont {Xue}, \citenamefont {Wang}, \citenamefont {Zhang},\ and\
  \citenamefont {Hu}}]{du2015nonadiabatic}%
  \BibitemOpen
  \bibfield  {author} {\bibinfo {author} {\bibfnamefont {H.}~\bibnamefont
  {Du}}, \bibinfo {author} {\bibfnamefont {S.}~\bibnamefont {Xue}}, \bibinfo
  {author} {\bibfnamefont {H.}~\bibnamefont {Wang}}, \bibinfo {author}
  {\bibfnamefont {Z.}~\bibnamefont {Zhang}}, \ and\ \bibinfo {author}
  {\bibfnamefont {B.}~\bibnamefont {Hu}},\ }\bibfield  {title} {\enquote
  {\bibinfo {title} {Nonadiabatic spectral redshift of high-order harmonics
  with the help of a vuv pulse},}\ }\href@noop {} {\bibfield  {journal}
  {\bibinfo  {journal} {Physical Review A}\ }\textbf {\bibinfo {volume} {91}},\
  \bibinfo {pages} {063844} (\bibinfo {year} {2015})}\BibitemShut {NoStop}%
\bibitem [{\citenamefont {Bian}\ and\ \citenamefont
  {Bandrauk}(2011)}]{bian2011nonadiabatic}%
  \BibitemOpen
  \bibfield  {author} {\bibinfo {author} {\bibfnamefont {X.-B.}\ \bibnamefont
  {Bian}}\ and\ \bibinfo {author} {\bibfnamefont {A.~D.}\ \bibnamefont
  {Bandrauk}},\ }\bibfield  {title} {\enquote {\bibinfo {title} {Nonadiabatic
  molecular high-order harmonic generation from polar molecules: Spectral
  redshift},}\ }\href@noop {} {\bibfield  {journal} {\bibinfo  {journal}
  {Physical Review A}\ }\textbf {\bibinfo {volume} {83}},\ \bibinfo {pages}
  {041403} (\bibinfo {year} {2011})}\BibitemShut {NoStop}%
\bibitem [{\citenamefont {Bian}\ and\ \citenamefont
  {Bandrauk}(2014)}]{bian2014probing}%
  \BibitemOpen
  \bibfield  {author} {\bibinfo {author} {\bibfnamefont {X.-B.}\ \bibnamefont
  {Bian}}\ and\ \bibinfo {author} {\bibfnamefont {A.~D.}\ \bibnamefont
  {Bandrauk}},\ }\bibfield  {title} {\enquote {\bibinfo {title} {Probing
  nuclear motion by frequency modulation of molecular high-order harmonic
  generation},}\ }\href@noop {} {\bibfield  {journal} {\bibinfo  {journal}
  {Physical review letters}\ }\textbf {\bibinfo {volume} {113}},\ \bibinfo
  {pages} {193901} (\bibinfo {year} {2014})}\BibitemShut {NoStop}%
\bibitem [{\citenamefont {Jia}, \citenamefont {Huang},\ and\ \citenamefont
  {Bian}(2017)}]{jia2017nonadiabatic}%
  \BibitemOpen
  \bibfield  {author} {\bibinfo {author} {\bibfnamefont {G.-R.}\ \bibnamefont
  {Jia}}, \bibinfo {author} {\bibfnamefont {X.-H.}\ \bibnamefont {Huang}}, \
  and\ \bibinfo {author} {\bibfnamefont {X.-B.}\ \bibnamefont {Bian}},\
  }\bibfield  {title} {\enquote {\bibinfo {title} {Nonadiabatic redshifts in
  high-order harmonic generation from solids},}\ }\href@noop {} {\bibfield
  {journal} {\bibinfo  {journal} {Optics Express}\ }\textbf {\bibinfo {volume}
  {25}},\ \bibinfo {pages} {23654--23662} (\bibinfo {year} {2017})}\BibitemShut
  {NoStop}%
\bibitem [{\citenamefont {Kim}\ \emph {et~al.}(2019)\citenamefont {Kim},
  \citenamefont {Shao}, \citenamefont {Kim}, \citenamefont {Han}, \citenamefont
  {Kim}, \citenamefont {Ciappina}, \citenamefont {Bian},\ and\ \citenamefont
  {Kim}}]{kim2019spectral}%
  \BibitemOpen
  \bibfield  {author} {\bibinfo {author} {\bibfnamefont {Y.~W.}\ \bibnamefont
  {Kim}}, \bibinfo {author} {\bibfnamefont {T.-J.}\ \bibnamefont {Shao}},
  \bibinfo {author} {\bibfnamefont {H.}~\bibnamefont {Kim}}, \bibinfo {author}
  {\bibfnamefont {S.}~\bibnamefont {Han}}, \bibinfo {author} {\bibfnamefont
  {S.}~\bibnamefont {Kim}}, \bibinfo {author} {\bibfnamefont {M.}~\bibnamefont
  {Ciappina}}, \bibinfo {author} {\bibfnamefont {X.-B.}\ \bibnamefont {Bian}},
  \ and\ \bibinfo {author} {\bibfnamefont {S.-W.}\ \bibnamefont {Kim}},\
  }\bibfield  {title} {\enquote {\bibinfo {title} {Spectral interference in
  high harmonic generation from solids},}\ }\href@noop {} {\bibfield  {journal}
  {\bibinfo  {journal} {ACS Photonics}\ }\textbf {\bibinfo {volume} {6}},\
  \bibinfo {pages} {851--857} (\bibinfo {year} {2019})}\BibitemShut {NoStop}%
\bibitem [{\citenamefont {Bandres}\ and\ \citenamefont
  {Guti{\'e}rrez-Vega}(2004)}]{bandres2004ince}%
  \BibitemOpen
  \bibfield  {author} {\bibinfo {author} {\bibfnamefont {M.~A.}\ \bibnamefont
  {Bandres}}\ and\ \bibinfo {author} {\bibfnamefont {J.~C.}\ \bibnamefont
  {Guti{\'e}rrez-Vega}},\ }\bibfield  {title} {\enquote {\bibinfo {title}
  {Ince--gaussian beams},}\ }\href@noop {} {\bibfield  {journal} {\bibinfo
  {journal} {Optics letters}\ }\textbf {\bibinfo {volume} {29}},\ \bibinfo
  {pages} {144--146} (\bibinfo {year} {2004})}\BibitemShut {NoStop}%
\bibitem [{\citenamefont {Couairon}\ \emph {et~al.}(2005)\citenamefont
  {Couairon}, \citenamefont {Sudrie}, \citenamefont {Franco}, \citenamefont
  {Prade},\ and\ \citenamefont {Mysyrowicz}}]{couairon2005filamentation}%
  \BibitemOpen
  \bibfield  {author} {\bibinfo {author} {\bibfnamefont {A.}~\bibnamefont
  {Couairon}}, \bibinfo {author} {\bibfnamefont {L.}~\bibnamefont {Sudrie}},
  \bibinfo {author} {\bibfnamefont {M.}~\bibnamefont {Franco}}, \bibinfo
  {author} {\bibfnamefont {B.}~\bibnamefont {Prade}}, \ and\ \bibinfo {author}
  {\bibfnamefont {A.}~\bibnamefont {Mysyrowicz}},\ }\bibfield  {title}
  {\enquote {\bibinfo {title} {Filamentation and damage in fused silica induced
  by tightly focused femtosecond laser pulses},}\ }\href@noop {} {\bibfield
  {journal} {\bibinfo  {journal} {Physical Review B}\ }\textbf {\bibinfo
  {volume} {71}},\ \bibinfo {pages} {125435} (\bibinfo {year}
  {2005})}\BibitemShut {NoStop}%
\bibitem [{\citenamefont {Franz}(2018)}]{franz2018high}%
  \BibitemOpen
  \bibfield  {author} {\bibinfo {author} {\bibfnamefont {D.}~\bibnamefont
  {Franz}},\ }\emph {\bibinfo {title} {High harmonic generation in crystals
  assisted by local field enhancement in nanostructures}},\ \href@noop {}
  {Ph.D. thesis},\ \bibinfo  {school} {Paris Saclay} (\bibinfo {year}
  {2018})\BibitemShut {NoStop}%
\end{thebibliography}%


\providecommand{\noopsort}[1]{}\providecommand{\singleletter}[1]{#1}%
%

\end{document}